\documentclass[fleqn,usenatbib]{mnras}

\usepackage{newtxtext,newtxmath}

\usepackage[T1]{fontenc}

\DeclareRobustCommand{\VAN}[3]{#2}
\let\VANthebibliography\thebibliography
\def\thebibliography{\DeclareRobustCommand{\VAN}[3]{##3}\VANthebibliography}

\usepackage{graphicx}	
\usepackage{amsmath}	
\usepackage{hyperref}
\usepackage{orcidlink} 
\usepackage{lineno} 

\defcitealias{Rotac}{ROTAC}

\title[Combining Rubin and Roman photometry]{Combining Rubin and Roman photometry and its implications for photometric redshift estimation}

\author[T. Shin et al.]{
Tae-hyeon~Shin$^{1}$\thanks{E-mail: taehyeos@andrew.cmu.edu}\,\orcidlink{0000-0002-6389-5409},
Rachel~Mandelbaum$^{1}$\,\orcidlink{0000-0003-2271-1527},
Chun-Hao~To$^{2,3,4}$\,\orcidlink{0000-0001-7836-2261},
Axel~Guinot$^{1}$\,\orcidlink{0000-0002-5068-7918},
Andrew~Hearin$^{5}$\,\orcidlink{0000-0003-2219-6852},
\newauthor
Arun~Kannawadi$^{6}$\,\orcidlink{0000-0001-8783-6529},
Chihway~Chang$^{2,3,4}$\,\orcidlink{0000-0002-7887-0896},
Federico~Berlfein$^{1,7}$\,\orcidlink{0009-0004-4113-9938},
Kaili~Cao$^{7,8,9}$\,\orcidlink{0000-0002-1699-6944},
Nihar~Dalal$^{7,8}$\,\orcidlink{0000-0003-2177-9407},
\newauthor
Michael~Gabe$^{7,8}$\,\orcidlink{0009-0006-0591-0080},
Christopher~Hirata$^{7,8}$\,\orcidlink{0000-0002-2951-4932},
Katherine~Laliotis$^{7,8,10}$\,\orcidlink{0000-0002-6111-6061},
Sidney~Mau$^{6}$\,\orcidlink{0000-0003-3519-4004},
Michael~A.~Troxel$^{6}$\,\orcidlink{0000-0002-5622-5212},
\newauthor
Masaya~Yamamoto$^{11}$\,\orcidlink{0000-0003-1585-997},
for the Roman HLIS Cosmology Project Infrastructure Team
\\
$^{1}$ McWilliams Center for Cosmology and Astrophysics, Department of Physics, Carnegie Mellon University, Pittsburgh, PA, USA\\
$^{2}$ Department of Astronomy and Astrophysics, The University of Chicago, Chicago, IL 60637, USA\\
$^{3}$ Kavli Institute for Cosmological Physics, University of Chicago, Chicago, IL 60637, USA\\
$^{4}$ NSF-Simons AI Institute for the Sky (SkAI),172 E. Chestnut St., Chicago, IL 60611, USA\\
$^{5}$ HEP Division, Argonne National Laboratory, 9700 South Cass Avenue, Lemont, IL 60439, USA\\
$^{6}$ Department of Physics, Duke University, Durham, NC 27708, USA \\
$^{7}$ Center for Cosmology and AstroParticle Physics (CCAPP), The Ohio State University, 191 West Woodruff Avenue, Columbus, OH 43210, USA \\
$^{8}$ Department of Physics, The Ohio State University, 191 West Woodruff Avenue, Columbus, OH 43210, USA \\
$^{9}$ Michigan Institute for Data and AI in Society (MIDAS), University of Michigan, 500 Church Street, Ann Arbor, MI, 48109, USA \\
$^{10}$ Université Paris Cité, CNRS, CEA, AstroParticule et Cosmologie, F-75013 Paris, France
}

\date{Accepted XXX. Received YYY; in original form ZZZ}

\pubyear{\the\year{}}

\begin{document}
\label{firstpage}
\pagerange{\pageref{firstpage}--\pageref{lastpage}}
\maketitle

\begin{abstract}
We perform a joint Roman--Rubin photometry analysis using the LSST Science Pipelines (LSP) on simulated images, by using the Roman F158 band as the reference band and applying forced photometry on the other bands. We find that the LSP CModel photometry for Roman bands is on average unbiased, but with a bright-end systematic uncertainty floor of 7-8\%. The \texttt{pgauss} photometry as part of the \texttt{Metadetection} weak-lensing shape measurement pipeline also exhibits a similar floor despite a substantially different flux-bias behavior. However, this systematic uncertainty floor becomes negligible for the color residuals and both algorithms recover unbiased colors for Roman as well as Rubin. We also show that not accounting for correlated pixel noise in Roman \texttt{IMCOM} coadded images results in underestimation of flux errors by a factor of $\sim$4. Based on the measured photometry, we build a machine-learning algorithm that maps the true galaxy properties to empirical flux errors, providing a convenient method to reproduce realistic photometry errors without having to rerun expensive image simulations and photometry pipelines. We apply this empirical error model to a new mock galaxy catalog with improved realism to investigate the impact of different survey depths and configurations on photometric redshift (photo-$z$) estimation performance. We find that 1) Rubin survey depth significantly impacts the photo-$z$ precision, 2) additional systematic uncertainties on photometry could significantly degrade the photo-$z$ performance especially towards high redshifts, therefore eroding the Roman--Rubin synergy, and 3) Roman photometry becomes important above $z\sim1.5$ while adding Rubin photometry to that of Roman improves photo-$z$ performance across all redshifts. Our results provide a foundation for understanding how realistic uncertainties in galaxy photometry may propagate into the Roman--Rubin synergy in photo-$z$ analyses.

\end{abstract}




\section{Introduction}
\label{sec:introduction}

The Nancy Grace Roman Space Telescope will soon begin its 5-year mission, which includes the High-Latitude Wide-Area Survey \citep[HLWAS;][ROTAC hereafter]{Rotac} among the three Core Community Surveys. The HLWAS is a multi-tiered imaging and spectroscopic survey, the core of which is the medium tier ($\sim$70\% of the allocated observing time), which will observe $\sim$2,300 deg$^2$ in the F106, F129, and F158 bands, in addition to grism spectroscopy. The depth per band is expected to reach an AB magnitude of $\sim$26.5 for point sources. Combined with the wide tier ($\sim$15\% of the allocated observing time, $\sim$2,200 deg$^2$ in the single F158 band), it will cover $\sim$4,500 deg$^2$ of the sky (Roman HLIS Cosmology PIT et al., in prepatation).\footnote{Note that these allocations and sky coverage are subject to possible changes.} The main mission of the HLWAS is to probe the origin of cosmic acceleration and the nature of dark matter using weak lensing, baryon acoustic oscillations, redshift-space distortion, galaxy clustering, and strong lensing. With its combination of wide-area multi-band imaging and grism spectroscopy, the HLWAS medium tier will form the primary observational basis for many of Roman's cosmological analyses.

The combination of weak lensing and galaxy clustering forms a powerful probe of cosmological parameters. This so-called ``3$\times$2-point'' analysis has been a backbone of modern cosmological analyses \citep[e.g.,][]{DESY6,Heymans2021,Zhang2026,Cao2026Fisher}. The abundance of galaxy clusters combined with the weak-lensing mass constraints also constitutes a strong cosmological probe \citep{Mants2015,Ghirardini2024,Bocquet2025,DESY3Cluster}.  \citet{Eifler2021,Eifler2021Roman} forecasts the constraining power of multi-probe cosmology (3$\times$2-point, cluster abundance, supernovae, redshift-space distortion, baryon acoustic oscillation) for the Roman High-Latitude Surveys, both alone and in combination with Rubin data. However, these analyses of photometric survey data rely on accurate redshift estimates of the lenses (foreground galaxies/clusters that cause gravitational lensing) and sources (background galaxies distorted by gravitational lensing). Therefore, accurate estimation of the photometric redshifts (photo-$z$s) of galaxies must precede the subsequent cosmological analyses \citep{Huterer2006,Ma2006,Lima2007,Mandelbaum2008,Nakajima2012,Hoekstra2015,vandenBusch2022,Rau2023,SOMPZ_DESY3,Wright2025}. 

There are two main methodological branches of photo-$z$ estimation: template-fitting methods using spectral energy distribution (SED) templates, and machine-learning methods. The template-fitting methods include, for example, \texttt{BPZ} \citep{BPZ}, \texttt{EAZY} \citep{EAZY}, and \texttt{LePHARE} \citep{Arnouts1999,Ilbert2006}. These methods share a similar photo-$z$ estimation logic: using SED templates, finding the best fit to the observed photometry. On the other hand, many different machine-learning-based photo-$z$ algorithms have been proposed \citep[e.g.,][]{DNF,FZBoost2,Graham2018}. For cosmological weak lensing and 3$\times$2-point analysis, point estimates of photo-$z$ are commonly used to divide galaxy samples into tomographic bins. While the aforementioned methods focus on deriving the point estimates of photo-$z$s and the corresponding probability distributions, $p(z)$, the \texttt{SOMPZ} method \citep[e.g.,][]{CarrascoKind2014,Masters2017,Hemmati2019} uses self-organizing maps (SOMs) based on galaxy colors to assign each galaxy into a low-dimensional ``cell'' associated with a certain redshift distribution rather than a point estimate. We refer readers to the references above for details of each proposed algorithm, and to \citet{Salvato2019} and \citet{Newman2022} for comprehensive reviews of photo-$z$ estimation methods. Most of these methods are wrapped as part of the \texttt{RAIL} package \citep{RAIL}, which we use in this study.

Using this photo-$z$ information, we can constrain the ensemble redshift distribution, $n(z)$, for each tomographic bin, a key ingredient in 3$\times$2-point cosmological analyses. There are several approaches to obtain $n(z)$: weighted stacking or stacking with spectroscopic calibration \citep{Bordoloi2010,Euclid2021}, a hierarchical Bayesian formalism \citep{Leistedt2016,Rau2023} for joint $p(z)$--$n(z)$ inference, and the \texttt{SOMPZ} method \citep{Myles2021,SOMPZ_DESY3,SOMPZ_DESY6}. For a more robust estimate, these approaches can be combined with clustering redshifts, which constrain $n(z)$ through spatial cross-correlations with spectroscopic samples \citep{Newman2008,Sanchez2019,Rau2022,Rau2023}. \citet{To2026}, using the same simulated data as that of ours, investigated in detail the impact of the quality of photometry on $n(z)$ distribution. While it used a different photometry pipeline (\texttt{slimfarmer}) from what is used in this analysis, it provides information complementary to this study which focuses on individual $p(z)$.

Meanwhile, the quality of photometric redshift estimates is limited by, e.g., the quality and wavelength coverage of multi-band photometry, biases in measured galaxy photometry (colors), blending of objects, and mismatches in point-spread function (PSF) modeling. The most important factors among these are the quality of galaxy photometry or colors (observation depth, flux/color bias, etc.) and the wavelength coverage \citep{Benitez2009,Dahlen2013,Ascaso2015,Beck2016,Graham2020}. However, as the HLWAS medium tier only covers wavelengths from approximately 1 to 1.8 $\mu$m, it lacks information on the 4000 Å break (the most important feature for photo-$z$ estimation) for galaxies at redshifts below $\sim$1.5, for which optical information is needed.

The Vera C. Rubin Observatory's Legacy Survey of Space and Time \citep[LSST;][]{LSST} bridges this information gap by providing optical coverage from approximately 0.35 to 1.0 $\mu$m. Meanwhile, the high-resolution NIR imaging of Roman has been shown as promising to identify the majority of unrecognized blends (blends that are not separable by deblending algorithms) in Rubin observations \citep{Troxel2023,Adari2025}, improve centroid measurements, and enhance photo-$z$ performance at high redshifts ($z$$>$1.5). Therefore, by combining the Roman HLWAS with the LSST, one can exploit their complementary wavelength coverage and angular resolution to construct galaxy samples with more accurate photometry and more reliable redshift characterization for subsequent cosmological analyses than either survey can provide alone \citep{Hemmati2019,Graham2020,Eifler2021,Mitra2021,Newman2022,Troxel2023,Bianco2024,To2026}. 

However, as the photo-$z$ estimation utilizes measured galaxy colors, its performance depends heavily on the quality of the photometry \citep{Bordoloi2010,Margoniner2008,Hildebrandt2012,Beck2017,Soo2018,Graham2018,MacCrann2022,Hang2024,Merz2025,To2026}. In order to fully leverage the synergy between these two surveys, we need to first investigate how the photometry quality and survey specifications affect photo-$z$ performance and how to optimize the corresponding photometry pipeline and survey configuration. 

The LSST Science Pipelines \citep{LSP} provide such a software library. The LSST Science Pipelines are an open-source software suite designed primarily to process Rubin observations to produce calibrated images, coadds, and catalogs of detected and measured astronomical sources \citep[e.g.,][]{DP2}. Owing to their flexibility in processing optical and NIR imaging data, they can serve as a platform to jointly process Rubin and Roman images. Given their operational readiness for processing Rubin data, the LSST Science Pipelines provide a natural framework for jointly processing Rubin and Roman imaging and for assessing how different photometry and survey configurations affect photo-$z$ performance while holding the detection, deblending, and flux-measurement methods fixed. Therefore, in this study, we focus on testing the photometry performance of the LSST Science Pipelines in processing combined Rubin and Roman data and its impact on the resultant photo-$z$ estimation under different survey and photometry configurations, using the simulated Rubin and Roman images from the OpenUniverse2024 simulation suite \citep{OU2024}. 

We additionally consider photometry measured as part of the \texttt{Metadetection} pipeline \citep{Sheldon2023}, a state-of-the-art shear-measurement framework used to construct the Dark Energy Survey Year-6 weak-lensing source catalog~\citep{Yamamoto2025}, Rubin data preview 2~\citep{DP2}  and likely to be adopted by the future weak-lensing surveys. Because the photo-$z$s of weak-lensing sources depend on the photometry associated with the shear-selected sample and a source detection process is part of the shear measurement pipeline in \texttt{Metadetection}\footnote{In standard \texttt{Metacalibration} \citep{Sheldon2017,Huff2017}, objects are detected before the artificial shearing operation, whereas \texttt{Metadetection} repeats the detection on the sheared images to account for shear-dependent selection biases.}, we explicitly compare the \texttt{Metadetection} photometry, implemented with \texttt{ngmix}\footnote{https://github.com/esheldon/ngmix} \citep{ngmix}, with that from the LSST Science Pipelines using the same data set, providing a controlled comparison of photometry between the two frameworks.

One of the main goals in this study is to further identify key sources of uncertainty in magnitude and color measurement. Using isolated galaxies without prominent blended neighbors, we assess how much object blending affects the Roman photometry performance. In addition, to cleanly separate systematic uncertainties from the total uncertainty budget, we calculate statistical flux errors accounting for the significant noise correlation in Roman coadded images \citep{Yamamoto2024}.

When validating photometry with simulations using different input galaxies with different truth properties, a common bottleneck is the need to run expensive image simulations and the corresponding photometry pipeline for each input truth. In order to overcome this computational bottleneck, we develop a machine-learning-based flux error model that maps the truth galaxy properties to observed flux errors, which enables an efficient way to reproduce realistic photometry errors without having to rerun image simulations and photometry pipelines on a large data set, as long as the instrumental specifications remain identical. We apply this model to a newer version of the mock galaxy catalog which has improved realism compared with the one used in the OpenUniverse2024 simulations for our photo-$z$ analysis.

The central goal of this study is to assess the photometry performance of the LSST Science Pipelines using combined Rubin and Roman images, and to investigate the effect of different survey and photometry configurations on the photo-$z$ estimation performance. To support the analysis, we also compare the photometry from the LSST Science Pipelines with that from the \texttt{Metadetection} pipeline using the same Roman data, and examine the contributions of blending and correlated noise in the Roman images to magnitude and color errors. We then use a machine-learning-based model for photometry errors to carry the measured flux errors and their inter-band correlations into a more realistic mock galaxy catalog for the subsequent photo-$z$ analysis, without rerunning image simulations and photometry pipelines. Note that, in this work, we perform a comparative study on the impact of photometry performance of the LSST Science Pipelines on the Roman--Rubin photo-$z$ estimation under different survey configurations. While the mock galaxy catalog used in this study is not 100\% realistic, it is sufficient for a controlled comparison of the photometry performance and its impact on the photo-$z$ estimation given reasonable realism of input galaxies.

Our paper is organized as follows. In Sec.~\ref{sec:data}, we describe the simulated data sets we use in this analysis. In Sec.~\ref{sec:photometry}, we explain the details of the photometry pipeline we adopted and show the corresponding results. In Sec.~\ref{sec:error_modeling}, we describe how we use machine-learning methods to model the observed flux errors and the corresponding inter-band error correlations to apply to another mock galaxy catalog. Sec.~\ref{sec:photoz}, we demonstrate our main photo-$z$ results under different survey and photometry configurations. Finally, in Sec.~\ref{sec:conclusion}, we summarize our findings and discuss the remaining caveats. Throughout the paper, for clarity, we use Y, J, H, F and K to represent the F106, F129, F158, F184, and F213 filters respectively.

\section{Simulated data}
\label{sec:data}

In this section, we describe the simulated image and catalog data that we use to validate our photometry pipeline and photo-$z$ estimation.

\subsection{OpenUniverse 2024: Joint Rubin and Roman image simulations}
\label{subsec:OU2024}

Our analysis is based on the OpenUniverse 2024 (OU2024) simulation suites, which provide simulated Rubin and Roman images and catalogs with many realistic features. Here, we briefly summarize the simulation features relevant to this study, while referring readers to \citet{OU2024} and the references therein for more detail.

The mock galaxy sample used for simulating images is constructed using the GalSampler technique \citep{Hearin2020} to populate halos with central and satellite galaxies in the Outer Rim simulation \citep{OuterRim2019}. Starting with the Small MultiDark Planck (SMDPL) N-body simulation \citep[][$L_{\rm box} = 400~h^{-1} {\rm Mpc}$]{MultiDarkSimulation} with star formation histories (SFHs) painted using the UniverseMachine \citep{UniverseMachine}, it substitutes these SFHs with parametric approximations using \texttt{Diffstar} and \texttt{Diffmah} models \citep{Diffstar, Diffmah}. After adding synthetic galaxies that lie below the simulation mass resolution limit with the same models, it transfers this galaxy population to the target simulation (Outer Rim in this case). The spectral energy distributions (SEDs) and photometry are calculated from the pre-computed SFHs, including metallicity, galaxy dust attenuation, and burstiness. Also, each galaxy is associated with a composite bulge, disk and star-forming knots profile, with the ellipticity assigned according to \citet{cosmoDC2}. This prescription, as implemented in the \texttt{Diffsky} package\footnote{https://github.com/ArgonneCPAC/diffsky}, results in a realistic galaxy population compared to real observations, with limitations such as the near-infrared (NIR) colors exhibiting narrower distributions than found in observations \citep{OU2024,To2026}.

\begin{figure*}
	\includegraphics[width=\textwidth]{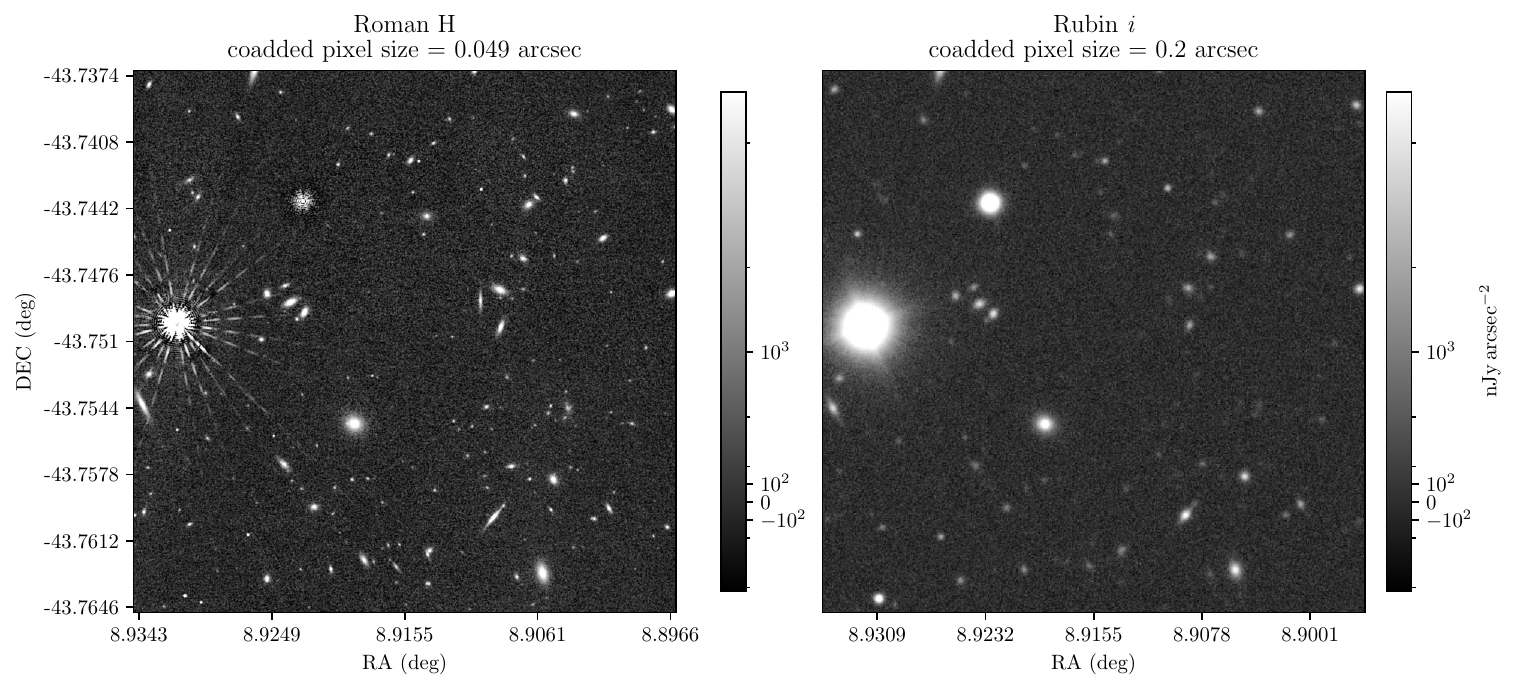}
    \caption{A randomly selected $100\arcsec\times100\arcsec$ cutout field for the Roman H band IMCOM coadd (Left) and Rubin $i$ band coadd. The Roman image corresponds to the nominal depth of the Roman medium HLWAS tier, and the Rubin image to the first five years of the LSST (as of August 2023; OpSim v3.2) with a caveat that the single exposures erroneously have 15s of exposure time rather than expected 30s ($\sim$0.37 magnitude shallower than the expected LSST-Y5 depth). For the image normalization, we first convert the pixel values to the unit of nJy, and use an arcsinh scale with the minimum of 1st percentile and the maximum of 99.5th percentile of the pixel values, with a linear threshold of 10\% of the difference between the minimum and the maximum. 
    }
    \label{fig:example_images}
\end{figure*}

OU2024 adopted the Rubin LSST observation sequence from Operations Simulator (OpSim) v3.2 \citep[Aug 2023,][]{Delgado2014,Delgado2016,Reuter2016}, spanning the first five of the ten-year rolling cadence (note that the observing strategy has evolved substantially since then). The images are simulated using the GalSim-based\footnote{https://github.com/GalSim-developers/GalSim} \texttt{imSim} framework \citep{DC2}\footnote{https://github.com/LSSTDESC/imSim} with updates in the realism including optical ray tracing, sensor/optics throughputs, airmass-dependent atmospheric transmission and expanded control of the telescope optical system. However, due to an error in interpretation of the simulated observing sequence, the output Rubin images involve a caveat relevant to this study: each exposure was only 15~seconds long, rather than the intended 30~seconds. This resulted in a depth loss of $\sim$0.37 magnitude. The images were processed and coadded using the LSST Science Pipelines \citep{LSP,Bosch2018,Bosch2019} with the data handled through the Rubin Data Butler \citep{Butler}. We use the coadded Rubin images in all six bands (\textit{ugrizy}), which have a pixel scale of 0.2$\arcsec$ and is accompanied by PSF models, World Coordinate System (WCS) solutions, and variance maps. In the right panel of Fig.~\ref{fig:example_images}, we plot an example $100\arcsec \times 100\arcsec$ cutout of a Rubin coadded image.

The corresponding Roman single exposure and coadded images were created according to the HLWAS observing strategy available at the time. However, given the evolved observing strategy for HLWAS and the simulation tools since the publication of the data, we use newer versions of simulated Roman HLWAS coadded images, which we will describe in the next section.

\subsection{Roman image coaddition with \texttt{IMCOM}}
\label{subsec:RomanIMCOM}

For Roman coadded images, we use the new \texttt{DC25} simulation suites (Roman HLIS Cosmology PIT et al., in preparation). The \texttt{DC25} simulation uses the same truth mock galaxy catalog described in the previous section, which is then used to generate single-visit images with various detector effects implemented in \texttt{romanisim}\footnote{https://github.com/spacetelescope/romanisim} as wrapped in \texttt{romanimpreprocess}\footnote{https://github.com/Roman-HLIS-Cosmology-PIT/romanimpreprocess}. These images are then processed through the pipeline developed by the Roman High-Latitude Imaging Survey Project Infrastructure Team (HLWAS; Roman HLIS Cosmology PIT et al., in preparation), following the up-to-date Roman HLWAS medium-tier observing strategy \citepalias[e.g.,][]{Rotac}. The final coadded images comprise four ``mosaics'', each of which covers $\sim$1.23 deg$^2$ of the sky in Y, J, H, F, and K (with the exception that the F and K bands cover only one mosaic). In this analysis, we only use one of the available mosaics to balance between enough statistical power for photometry and computation time. 

The image processing is divided into four stages:
\begin{itemize}
    \item Level 1, uncalibrated raw data: raw images including various detector and physical effects.
    \item Level 2, calibrated images: calibrations including reference pixel correction, dark-current correction, flat-field correction, bias correction, classical linearity, inter-pixel capacitance, ramp fitting \citep{Casertano2022}, jump detection \citep{Sharma2024}, and sky subtraction. 
    \item Level 2.1, destriped and masked images: image destriping \citep{imdestrip} and outlier flagging from difference images.
    \item Level 3, coadded images: image coaddition carried out using \texttt{pyimcom} \citep{Rowe2011,Hirata2024,Yamamoto2024,Cao2025}, a linear-algebra-based image combination algorithm to achieve an oversampled coadded PSF from undersampled images.
\end{itemize}

We use the coadded images in the Y, J, and H bands with an output Gaussian PSF \citep{Cao2026} with FWHM (full width at half maximum) of 0.24$\arcsec$ and the pixel size of $s_{\rm pix,imcom}= 0.049\arcsec$. Following the current plan for the Roman HLWAS medium-tier survey \citepalias{Rotac}, we only use these three bands for our Roman analysis. We show an example $100\arcsec \times 100\arcsec$ cutout of a Roman \texttt{IMCOM} coadded image in the left panel of Fig.~\ref{fig:example_images}. 

Each coadded image is associated with additional image layers including the truth image layer, four realizations of the noise field with only background and instrumental noise, and another four realizations of the noise field additionally including self-Poisson noise from astronomical sources. These noise fields are processed through the same pipeline described above. We refer readers to \citet{To2026} and the upcoming Roman HLIS Cosmology PIT et al. (in preparation) for more detail. Note that due to an unresolved issue in background subtraction (which is beyond the scope of this paper), we use the coadded images constructed by adding the truth images to the noise layers including the self-Poisson noise (see Sec.~\ref{subsec:photometry_lsp}).

\subsection{Updated \texttt{Diffsky} mock galaxy catalog}
\label{subsec:diffsky}

As described in Sec.~\ref{subsec:OU2024}, \texttt{Diffsky} \citep{hearin_etal23_dsps,alarcon_etal25_diffstarpop} is a model that connects the SED of galaxies with the mass assembly history (MAH) of their underlying dark matter halos. An early prototype of \texttt{Diffsky} was used to create the OU2024 mocks in \citet{OU2024}, as described in the previous section. The updated mock galaxy catalog was instead created by directly populating the merger trees in the Last Journey simulation \citep{heitmann_etal21_last_journey} with an updated formulation of the \texttt{Diffsky} model, selecting best-fit parameters derived by fitting the photometry in the COSMOS-20 dataset \citep{cosmos20_weaver_etal}. 

The particular $N$-body data products used by \texttt{Diffsky} to create synthetic galaxy catalogs are the merger trees of {\tt subhalo cores}, a computationally efficient alternative to conventional subhalo tracking algorithms~\citep{2023OJAp....6E..24K, sultan_etal21_last_journey2}. The Last Journey simulation resolves cores down to a halo mass of $M_{\rm halo}\gtrsim 10^{11.5}M_{\odot};$ galaxies in the updated catalog additionally populate a fully synthetic halo population in the mass range $10^{10.5}<M_{\rm halo}/M_{\odot}<10^{11.5}$ using \texttt{Diffhalos} \citep{zacharegkas_etal26_diffhalos} to generate the synthetic halos. For photo-$z$ analyses, we use galaxies created by this updated \texttt{Diffsky} algorithm with empirical and theoretical photometry errors assigned according to our models described in Sec.~\ref{sec:error_modeling}.

\subsection{The Rubin Operations Simulator}
\label{subsec:opsim}

In OU2024, the v3.2 of the Rubin operational simulations\footnote{https://community.lsst.org/t/baseline-v3-2-released/7877}, as part of \texttt{rubin\_sim}\footnote{https://github.com/lsst/rubin\_sim}, was used to account for the LSST survey strategy to generate the output LSST images. However, with a series of major updates to the LSST survey strategy, multiple subsequent sets of operational simulations have been released. These updates include, for example, the updated rolling cadence that ensures uniform coverage every three years after the first-year operation, updated deep drilling field strategy, the correspondingly updated depth distribution, modified system throughput due to the triple-silver coating of the mirrors, the exposure structure (from $1\times30s$ for $u$ band and $2\times15s$ for $grizy$ band per visit to $1\times38s$ and $1\times30s$, respectively), and updated readout time as well as slew model\footnote{https://survey-strategy.lsst.io/baseline/changes.html}. We refer readers to \citet{SCOCphase2,SCOCphase3} and the resources mentioned in this section for more details. 

We use this updated version (v5.0.1) of the Rubin operational simulations\footnote{https://community.lsst.org/t/release-of-v5-0-simulations/11114} \citep{OpSimv5} to extract the survey depth, the number of visits per band ($N_{\rm visit}$), the noise properties, and the size of expected Rubin PSFs for our theoretical error calculation (Sec.~\ref{subsec:prepare_rubin_images} and \ref{subsec:error_theoretical}).\footnote{Note that currently there is the v5.3 runs, however, this does not significantly affect our results.}

To be specific, we first select the relevant Wide-Fast-Deep (WFD) footprint, removing the high-dust region near the Galactic center, Deep-Drilling Fields (DDF), Targets-of-Opportunity (ToOs), and twilight observations. We then extract $N_{\rm visit}$, sky brightness, and the PSF FWHM per each Healpix\footnote{https://healpix.sourceforge.io/} pixel with $N_{\rm side}=64$ \citep{Gorski2005,Zonca2019}. These quantities are then converted to the sky background and instrumental noise in the flux unit using the system zeropoint. Combined with the effective PSF area calculated from the PSF FWHM, we calculate the final flux noise assuming a point-source object by adding it to the self-Poisson flux noise. We refer readers to Sec.~\ref{subsec:prepare_rubin_images} and \ref{subsec:error_theoretical} for detailed implementation and correction for extended objects.

\section{Photometry}
\label{sec:photometry}

In this section, we describe the photometry algorithms adopted in this study and their performance along with their uncertainty estimates.

\subsection{Photometry with the LSST Science Pipelines}
\label{subsec:photometry_lsp}

The LSST Science Pipelines\footnote{https://pipelines.lsst.io} \citep[``LSP'' from now on;][]{LSP} are a suite of astronomical image and catalog data processing modules, designed primarily to process the data from the Rubin Observatory Legacy Survey of Space and Time (Rubin LSST). Due to its flexibility and comprehensive coverage from image preprocessing to multi-band photometry, it can be extended to process optical or near-infrared images from a range of telescopes. Moreover, the Rubin community has been already using the LSP to process Rubin observations \citep{DP2} and the Rubin photometry is crucial for the Roman photo-$z$. Therefore, in this study, we use the LSP to process both Roman and Rubin data, as described below. 

We choose the Roman H band as our reference band in which the reference photometry and the selection are made, motivated by the proposed Roman HLWAS design in which the H band provides the broadest imaging coverage, including both the medium tier (in Y, J, and H) and the H-only wide tier \citep{Rotac}. One known issue for the \texttt{DC25} simulated coadds is the over-subtraction of background \citep[$\sim$0.02 magnitude at $m_{\rm H} = 18$ and $\sim$0.13 magnitude at $m_{\rm H} = 24.5$;][]{To2026}. To separate the effect of the background subtraction from our main interests, we use simplified images as input for the photometry. More specifically, the observed images are constructed by adding a noise realization that includes background noise, instrumental noise, and source self-Poisson noise to the truth image, which were processed consistently through the \texttt{IMCOM} pipeline. Therefore it retains pertinent noise properties (Sec.~\ref{subsec:RomanIMCOM}) and does not suffer from imperfect background subtraction by construction. 

\textbf{Construction of variance plane.} For source detection, we use the mean image of the Y, J, and H bands to reduce noise and maximize the detection capabilities. Since the noise levels in the Y, J, and H band are within 5\% from each other, we use the simple mean rather than a weighted average. We start by constructing the variance plane. As derived in \citet{To2026}, the background variance field for a given coadded image $\alpha$ with contributing single exposures $i$ can be written as:
\begin{equation}
\label{eq:background_var}
    V_{\textrm{bkg},\alpha}
    = \sum_i \left(T^i_\alpha\right)^2
    \frac{\sum_{\alpha'} N_{\alpha'}^2}{\sum_{\alpha'} \sum_{i'} \left(T_{\alpha'}^{i'}\right)^2}, 
\end{equation}
where $N_\alpha$ is the background- and instrumental-only noise field (Sec.~\ref{subsec:RomanIMCOM}), and $T^i_\alpha$ the linear transformation matrix from the original single exposure image pixels ($i$) to the final \texttt{IMCOM} output coadd pixels ($\alpha$), provided as an additional output map as $(T^i_\alpha)^2$ (noise amplification). For clarity of notation, we omit $\alpha$ from the subsequent equations. Then the variance field used for the detection (the mean image) becomes\footnote{Note that we use this background-only variance field for the detection \citep{Sextractor,Sadr2019,Kelvin2023}. Including self-Poisson noise results in a detection threshold dependent on source fluxes. Also note that detection is a hypothesis test between background-only and background-plus-source, therefore the relevant variance is the background variance.},
\begin{equation}
    (V_{\rm bkg,Y} + V_{\rm bkg,J} + V_{\rm bkg,H})/9.
\end{equation}
Note that this variance calculation corresponds to a per-pixel variance, not including the correlation between adjacent pixels due to the \texttt{IMCOM} process \citep[correlation length $\sim$ output PSF size, and also dependent on the input PSF size; for more information, see Appendix A of][]{Yamamoto2024}. 

For the subsequent steps including deblending, flux measurements, and object selection, we add self-Poisson variance to the background variance. Since the image count has the unit of ${\rm DN}/(0.11 \arcsec)^2/s$ (where $s_{\rm pix,orig} = 0.11\arcsec$ is the pixel scale of the original Roman detector, and DN represents data number), the effective gain becomes:
\begin{equation}
    g_{\rm eff} = g_{\rm detector} \, t_{\rm exp} \, n_{\rm eff} \, s_{\rm pix,imcom}^2 / s_{\rm pix,orig}^2,
\end{equation}
where $g_{\rm detector} = 1.458$ is the detector gain, $t_{\rm exp}$ the exposure time for single exposures, and $n_{\rm eff}$ is the effective number of images used for coaddition. Then the self-Poisson variance field ($V_{\rm Poisson})$ can be constructed from the truth image layer ($I_{\rm truth}$) as:
\begin{equation}
    V_{\rm Poisson} = I_{\rm truth} / g_{\rm eff},
\end{equation}
so that the total variance is $V_{\rm total} = V_{\rm bkg} + V_{\rm Poisson}$.\footnote{For example, a point source with the H-band magnitude $\lesssim 22.8$ would have the self-Poisson contribution similar or greater than the background contribution. We refer readers to Appendix~C of \citet{To2026} for more detailed analysis.} 

\textbf{Source detection.} Using the LSP's \texttt{SourceDetectionTask} and the calculated per-pixel background variance as above, we detect sources that are above 4 times the standard deviation spanning at least 4 contiguous pixels (\texttt{thresholdValue}=4.0, \texttt{thresholdType}=``pixel\_stdev'', and \texttt{minPixels}=4 in the configuration). Note that, for the detection task here, we ignore the considerable pixel-to-pixel noise correlation (Sec.~\ref{subsec:correlated_noise}); instead, we apply a somewhat high detection threshold to avoid too many false detections while retaining the majority of the true objects. We do not convolve the image with the detection kernel as the \texttt{IMCOM} coadded images are effectively the sky scene convolved with the output PSF, and convolving with a detection kernel results in a significant fraction of false detections (the detection kernel amplifying the correlated noise peaks, which are approximately the size of the PSF itself). We then grow the source footprints isotropically by 1.5 times the PSF RMS width. The resultant detection catalog includes object IDs, footprints, and peak positions. 

\textbf{Deblending.} We then deblend the detected sources using \texttt{Scarlet} \citep{Scarlet}, as implemented in the LSP as \texttt{Scarlet Lite}\footnote{https://github.com/lsst/scarlet\_lite}. In \texttt{Scarlet} each source is represented by one or more nonnegative, radially monotonic components factorized by their shapes and wavelength-dependent amplitude in each band, initialized around the detected peaks. It then convolves the model with the image PSF and optimizes using a variance-weighted least-squares method. The final model is used to re-distribute the pixels among the deblended children and update the detection catalog accordingly. Since details of \texttt{Scarlet} and the corresponding impact on photometry are not one of our main interests, we do not use its multi-band deblending feature, and instead only use the mean image described above. We have checked that the specific configuration settings within reasonable ranges do not meaningfully alter the results. However, we do limit the maximum number of peaks at 30 (\texttt{maxNumberOfPeaks}=30) to reduce spurious deblending due to the correlated noise peaks around bright objects.

\textbf{Flux measurement.} This deblended detection catalog is passed to the measurement pipeline (\texttt{SingleFrameMeasurementTask})\footnote{Note that since we perform photometry on the H band and apply forced photometry on all other bands, a multi-band measurement is not needed.}. In addition to the default setting, we add two flux measurement algorithms which we use in this analysis: CModel \citep{Abazajian2004,Bosch2018} and GAaP \citep{Kuijken2008, 2022rubn.rept....3K} photometry implemented in LSP as \texttt{modelfit\_CModel} and \texttt{ext\_gaap\_GaapFlux}, respectively. CModel is a model-fitting flux measurement algorithm that uses a composite de Vaucouleurs (bulge) plus exponential (disk) profile\footnote{Note that this is almost identical to the input truth galaxies for our simulations, modulo knots that randomly follows the disk profile (Sec.~\ref{subsec:OU2024}).}. The algorithm implemented in the LSP follows that of the Sloan Digital Sky Survey \citep{Abazajian2004}. The GAaP (Gaussian-aperture-and-PSF) photometry \citep{Kuijken2008} is designed to measure galaxy colors accurately and consistently across bands with different seeing. It first Gaussianizes the PSF by convolving the image until the PSF becomes a slightly broader circular Gaussian (1.15$\times$ by default). Then it measures fluxes using a Gaussian aperture, chosen so that the effective pre-seeing Gaussian aperture is the same in every band. For the reference magnitude in H, we use CModel photometry, and for the color measurements we use the GAaP fluxes with the ``optimal'' Gaussian aperture where the aperture size and shape is determined from the measured second moments of the source image, convolved by the PSF (same for the Y and J bands).

\textbf{Forced photometry on the other Roman bands.} Given the final measurement catalog in the H band as the reference band, we perform forced photometry on the other two Roman bands (Y and J) using \texttt{ForcedMeasurementTask}. In this forced measurement pipeline, the galaxy ids, the source footprints, the centroid positions, and the shapes are re-used from the reference catalog. For CModel, also, the final composite ``shapes'' and fit region are recycled, so that it only optimizes the amplitudes. For other fields in the configuration, we use identical settings as for the H band.

\textbf{Forced photometry on the Rubin bands.} Finally, we perform forced photometry on the Rubin coadded images. Since the Roman and the Rubin coadded images have different pixel sizes (0.049$\arcsec$ for Roman and 0.2$\arcsec$ for Rubin), we first need to transform the detected source footprints, using the Rubin WCS and measured Rubin PSF. Given a Roman coadded image and the corresponding source catalog, we first identify the Rubin coadded images that include those source positions. We then transform the Roman footprints (and the corresponding source positions) into Rubin footprints using the Rubin WCS, and the transformed footprints are then dilated by 3$\sigma$ of the Rubin PSF to account for the change in PSF (with the Rubin PSF being significantly broader). These transformed and enlarged source footprints are then used for the forced photometry measurements. When we select sources to perform the forced photometry on the Rubin images, we only keep those that lie within the inner bounding box (excluding the buffer region) to keep unique sources without repetition. Note that, due to the current limitations of the forced photometry implemented in the LSP, model-fitting photometry such as CModel results in \texttt{NaN} fluxes if the pixel size is different from that of the reference band. Thus, we run CModel photometry only on Roman coadded images. Note that this does not cause any problem in our analysis, since our fiducial choice for galaxy colors is GAaP photometry.

\subsection{Preparing the Rubin images for measurement of GAaP photometry}
\label{subsec:prepare_rubin_images}

Before we perform the GAaP photometry on these Rubin forced sources, we first update the noise properties of the Rubin images to match the more recent LSST depth using the v5.0.1 of the Rubin operational simulations (Sec.~\ref{subsec:opsim}). Due to the shallower depth in our OU2024 Rubin coadded images than what is intended in the OpSim v3.2 (Sec.~\ref{subsec:OU2024}), the final depths of the Rubin coadded images are shallower than expected for year 4  in v5.0.1 (the first uniform coverage after Y1). Since we can only degrade and not improve the noise properties of images, we degrade the Rubin images in OU2024 to the LSST-Y1 depth from OpSim v5.0.1. We first obtain the current level of noise standard deviation by calculating the RMS of the pixels with the negative counts (assuming that the negative-count pixels are minimally affected by the sources), $\sigma_{\rm OU,Rubin} = \sqrt{V_{\rm OU,Rubin}}$ \citep{Heidt2003,Gorski2019}. 

Then following \citet{Jones2021}, we calculate the expected per-pixel background noise level from the OpSim database as:
\begin{equation}
    \sigma^2_{\rm pix} =\sigma^2_{\rm inst,pix} + \sigma^2_{\rm sky,pix},
\end{equation}
where,
\begin{equation}
    \sigma^2_{\rm inst,pix} = \frac{N_{\rm visit} \times (RN)^2 + I_{\rm dark} \times t_{\rm visit} \times N_{\rm visit}}{g},
\end{equation}
and
\begin{equation}
    \sigma^2_{\rm sky,pix} = \frac{F_{\rm sky,pix} }{g},
\end{equation}
where the gain g is assumed to be 1 in the entire Rubin simulations, $N_{\rm visit}$ is the number of visits, $RN=8.8$ is the read noise, $I_{\rm dark}=0.2$ is the dark current, $t_{\rm visit}$ is the exposure times per visit, $F_{\rm sky,pix}$ is the expected sky background count per pixel, and $\sigma_{\rm pix}$ represents the theoretical noise standard deviation per pixel. We calculate this $\sigma^2_{\rm pix}$ per HEALPix pixel of $N_{\rm side}=64$ and take the median value for our subsequent analysis. Note that we only select the wide-field area with the low dust region for our analysis. We degrade our Rubin coadded images to the LSST-Y1 depth using this calculated background noise level, by adding a Gaussian noise with the standard deviation of $\sqrt{V_{\rm pix} - V_{\rm OU,Rubin}}$ to the images\footnote{This is done after accounting for the difference in the magnitude zeropoint between the native Rubin detector and the coadded images in OU2024.}. The resultant images have the noise variance of $V_{\rm pix}$, our target noise level.

We then perform forced GAaP photometry on these degraded Rubin images, where we use the Gaussian kernel width of 1.0$\arcsec$ (due to the limitation in LSP, the ``optimal'' Gaussian apertures used for Roman images in Sec.~\ref{subsec:photometry_lsp} cannot be calculated). We confirmed that changing the kernel width within a reasonable range does not change our result qualitatively. Note that due to the limitation of the LSP, forced photometry on a different pixel size image cannot handle the ``optimal'' GAaP photometry.

\subsection{Photometry as part of a \texttt{Metadetection} pipeline}
\label{subsec:photometry_mdet}

\texttt{Metadetection} \citep{Sheldon2023,Yamamoto2025} is a weak-lensing shear measurement algorithm in which small artificial shears are applied to observed images, and the full detection, deblending, selection, and shape-measurement pipeline is run on each set of sheared images. By measuring how the final selected catalog responds to these known artificial shears, it estimates the total response of the ensemble of galaxy ellipticities with respect to the shear. It differs from \texttt{Metacalibration} \citep{Sheldon2017,Huff2017} in that detection and selection become part of the pipeline, naturally accounting for relevant selection effects, rather than being applied only to already detected sources. 

We run the \texttt{Metadetection} ``photometry'' algorithm on the Roman \texttt{IMCOM} coadded images in the same three bands. We then compare the quality of the photometry from \texttt{Metadetection}, implemented using the \texttt{ngmix} package \citep{ngmix}, to that of the LSP. 

For the detection, we use the \texttt{SEP} \citep{Barbary2016}\footnote{https://sep.readthedocs.io/en/stable/} package. \texttt{SEP} is an implementation of \texttt{Source Extractor} \citep{Sextractor}, written in the \texttt{Python} language. We use an absolute detection threshold of 0.02 pixel count, requiring 5 contiguous pixels for detection, using a Gaussian matched filter (kernel) with an FWHM of 0.24$\arcsec$ (the PSF size). 

The \texttt{Metadetection} flux measurement algorithm we adopt in this analysis is the \texttt{pgauss} algorithm. The \texttt{pgauss} measurements are estimates of pre-PSF Gaussian-weighted fluxes computed in Fourier space. For each detection, the method applies a fixed Gaussian aperture to the PSF-deconvolved galaxy image by multiplying the Fourier-transformed image by a stabilized kernel \(W(k)/P(k)\)\footnote{Here, $W(k)$ is the Gaussian weight function, and $P(k)$ is the Fourier-transformed PSF image.}, yielding matched-aperture fluxes across bands\footnote{The measured fluxes and colors are used for the object selection which is consistently propagated to the pipeline.}. Because its fixed pre-PSF Gaussian aperture matches across different bands, it is especially suitable for color measurements \citep{Yamamoto2025}. In this analysis, we use a Gaussian aperture with an FWHM of 0.5$\arcsec$. The flux errors are calculated from the variance map using the method described in \citet[][also see Sec.~\ref{subsec:photometry_lsp}]{To2026}. For details of the \texttt{Metadetection} photometry, we refer readers to the above references. Note that in this \texttt{Metadetection pgauss} photometry, we do not account for the noise correlation between pixels (Sec.~\ref{subsec:correlated_noise}, see Sec.~\ref{subsec:photometry_result_roman} for how we effectively match the sample from the LSP photometry to that of \texttt{Metadetection pgauss}.)

Note that the \texttt{pgauss} photometry described in this section does not include object deblending. Instead, we use the \texttt{überseg} algorithm \citep{Jarvis2016} where a pixel's weight is set to zero if 1) \texttt{Source Extractor} has assigned the pixel to another object, or 2) the pixel is closer to another detected object than the object of interest. Together with the Gaussian weight function applied as part of the algorithm, it minimizes the impact of neighbors to the measured photometry.

\subsection{Flux uncertainty with correlated noise}
\label{subsec:correlated_noise}

\begin{figure}
	\includegraphics[width=\columnwidth]{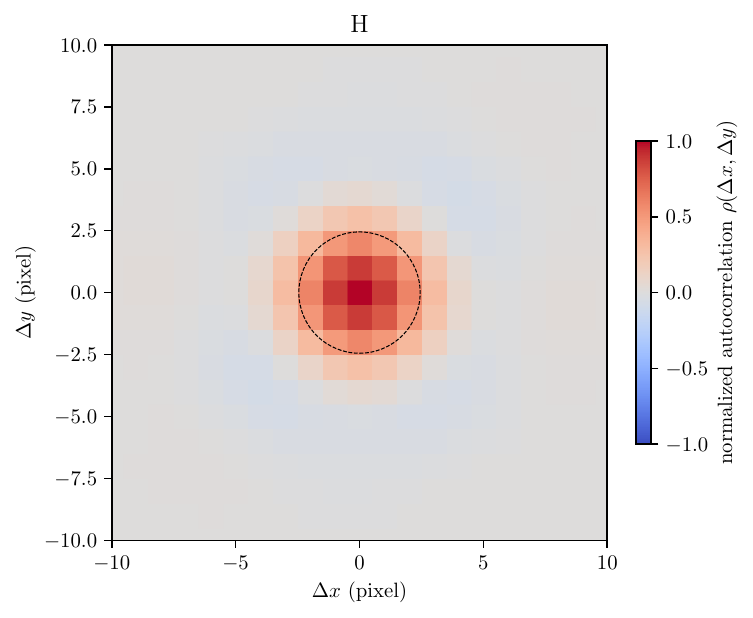}
    \caption{The normalized noise correlation functions for the H band. The dashed circle represents the PSF FWHM of 0.24$\arcsec$. We only show one band because the IMCOM output PSF is identical across the Y, J, and H bands, and the noise correlation structure is therefore similar across each band, despite the difference in the input PSF ($\lesssim 3$\% difference).}
    \label{fig:noise_correlation}
\end{figure}

Due to the transformation of the PSF from the original Roman PSF to the target Gaussian PSF with FWHM of 0.24$\arcsec$, the \texttt{IMCOM} coadded images have correlated noise with a correlation length that is approximately the size of the target PSF ($\sim$5 pixels). Without considering this correlated noise in the flux error calculation, the resultant flux error could be underestimated by factors of a few \citep{To2026}. Calculating correct flux uncertainties is important because 1) they impact SNR-based object selection which could propagate to downstream analyses, and 2) we need to understand the image noise properties to be able to precisely assess the photometry quality and potentially improve it.\footnote{Note that incorrect flux uncertainties can also, for example, affect the performance of red-sequence-based galaxy cluster finders \citep{Rykoff2014,Rykoff2016}, where the flux uncertainties are used to estimate the intrinsic width of the red-sequence in the color-magnitude space.} Fig.~\ref{fig:noise_correlation} shows the normalized noise correlation function retrieved from the background-only noise layers in one Roman band. The noise correlation indeed extends to $\sim$5 pixels as expected. In this section, we describe the formalism for calculating the flux errors with the correlated noise. 

We start by deciding the size of the cutout on which the errors are calculated. First we take the image second moments of the exponential component, the de Vaucouleurs component, and the composite profile from the catalog and from each second moment matrix calculate the eigenvalues, the largest of which corresponds to the square of the RMS major-axis size in pixels. Then we take the largest of the three and call it $r_{\rm max}$. Next, we define $h$ as,
\begin{equation}
   h = 12 r_{\rm max} + 3 {\rm FWHM}_{\rm PSF},
\end{equation}
clipped between 32 and 256. We use a $(2h+1) \times (2h+1)$ box size for the final cutout for the error calculation. Note that the first term in the equation corresponds to approximately $12\sigma$ for a Gaussian profile and fully accommodates the adopted CModel template truncation.

We then calculate the normalized correlation functions, $\rho (\Delta x, \Delta y)$, from the four background-only noise realizations as follows. We have checked that the effect of including the self-Poisson noise is dominated by a small fraction of extremely bright objects, therefore, we exclude it from the noise correlation calculation. We first subtract the means\footnote{This is to remove any remaining background component in the noise realizations. However, by construction, there is no background component in them, therefore it does not have any empirical effect on our analysis.} from the noise images then calculate the autocorrelation functions using the fast Fourier transform as implemented in \texttt{numpy}. Then we take the average of the four (to reduce noise), each of which is normalized so that $\rho(0, 0)=1$. After that, we take the fitted CModel templates of each source, $m_i$, normalized to the flux of unity, and the total variance, $V_{\rm tot} = V$ (Sec.~\ref{subsec:photometry_lsp}). Finally, the flux variance ($V_{\rm \hat{F}} = \sigma_{\rm \hat{F}}^2$) with correlated noise can be expressed as\footnote{This equation assumes the CModel fitting with an inverse-variance weighting.} \citep{To2026}: 
\begin{equation}
\label{eq:corr_error}
\begin{split}
    V_{\rm \hat{F},corr} &= \frac{
    \sum_i \sum_j \left[m_i m_j \rho(\boldsymbol{x}_i - \boldsymbol{x}_j) / \sqrt{V_i V_j}\right]
    }
    {
    \left[\sum_i \left( m_i^2 / V_i \right) \right]^2
    } \\
    &= \frac{\sum_i \sum_j \left[m_i m_j \rho(\boldsymbol{x}_i - \boldsymbol{x}_j) / \sqrt{V_i V_j}\right]}{
    \sum_i \left( m_i^2 / V_i \right)
    } \times V_{\rm \hat{F}, uncorr},
\end{split}
\end{equation}
where $V_{\rm \hat{F}, uncorr}$ is the flux variance without the correlated noise (diagonal) and the factor in front of it could be used as a correction factor. Here, $i$ and $j$ are pixel indices of the cutout.

For the Y, J, and H bands, the correction factors ($\sqrt{V_{\rm \hat{F},corr} / V_{\rm \hat{F},uncorr}}$) are calculated as $4.1^{+0.3}_{-0.2}$, $4.0^{+0.3}_{-0.2}$, and $4.0^{+0.1}_{-0.1}$, respectively (16th, 50th, and 84th percentile values from the sample). These correction factors depend weakly on galaxy magnitudes and sizes with $\sim$3-5\% variation between the lowest and highest quartiles. Therefore, without accounting for this correlated noise, the measured flux signal-to-noise ratio (SNR) would be overestimated by a factor of $\sim$4.

\subsection{Results for Roman photometry}
\label{subsec:photometry_result_roman}

To construct a ``clean'' SNR-limited sample free from detection systematics and mismatches to the truth, we impose the following selection criteria on our measured photometry catalog. Note that the investigation of false detections and other detection-related systematics is beyond the scope of this paper. 

We select the galaxy sample for our subsequent analysis steps with these criteria in the H band:
\begin{itemize}
    \item Flux SNR $>5$, using the correlated error, Eq.~\eqref{eq:corr_error}.
    \item Successful CModel fit.
    \item No deblended children.
    \item Mutually matched to the truth objects within 3 pixels ($\sim$0.15$\arcsec$). Here, ``mutually'' matched objects correspond to those matched mutually between the truth and the observed catalog.
    \item Magnitude difference to the matched truth object < 1.
    \item Object type = "galaxy" (using the matched truth information). 
\end{itemize}
The final number of selected galaxies in the H reference band is 273,451 ($\sim$$62/{\rm arcmin}^2$). We do not impose any additional selection to the Y and J bands, regardless of the validity of their flux measurements. Note that the mutual matching criterion and the cut on the magnitude difference are to select a clean galaxy sample, minimizing contamination from spurious matching. This helps us study the effects of the survey combination and depth without interference from detection systematics. Also note that we deliberately choose an SNR cut that is likely lower than will be used for cosmological analyses, so as to allow exploration of the impact of various selections on the results. 

\begin{figure}
	\includegraphics[width=\columnwidth]{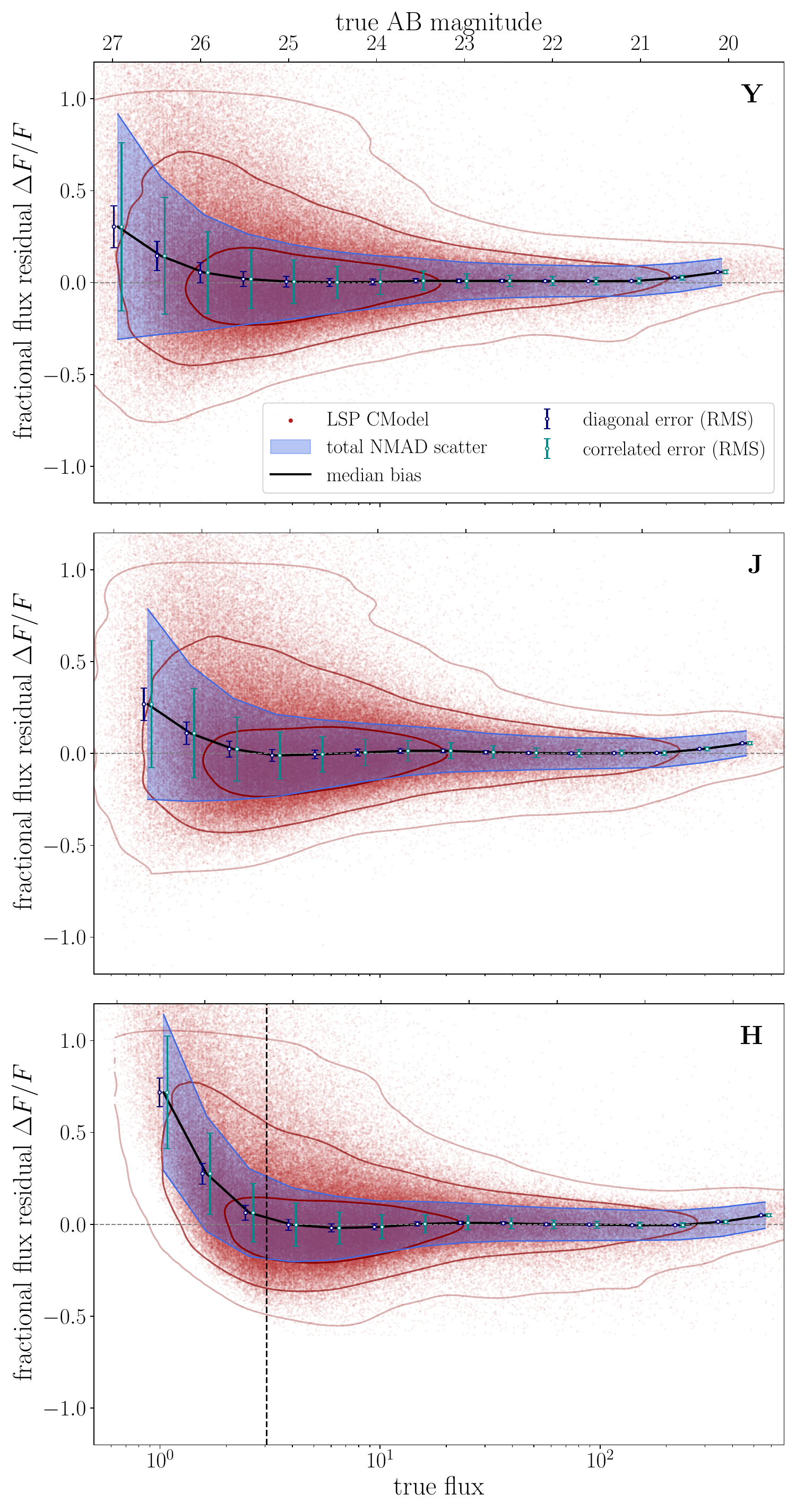}
    \caption{The fractional residuals of the measured fluxes using the LSST Science Pipelines CModel algorithm with respect to the true fluxes (red dots) as a function of true flux (lower axis) and magnitude (upper axis), for each of three Roman bands. Also plotted are the median flux bias (black line), the NMAD scatter of the measured residuals (blue shade), the RMS of the diagonal errors (navy error bars; without accounting for the noise correlation), and RMS of the correlated errors (cyan error bars; accounting for the noise correlation). We plot the Y, J and H bands from top to bottom. The contours represent the 1, 2, and 3$\sigma$ ranges in 2D. The vertical dashed line in the bottom panel represents the approximate 5$\sigma$ limiting magnitude expected \citepalias{Rotac}, which agrees well with our observation. Note that the flux bias is mostly consistent with zero, except for the expected selection effect on the dimmer end (especially for H, the reference band), and a slight upturn at the bright end (although not significant). Also note that the observed scatter (blue shades) reaches a systematic error floor at the bright end, rather than statistically asymptoting to zero.}
    \label{fig:flux_residual_lsp}
\end{figure}

\begin{figure}
    \includegraphics[width=\columnwidth]{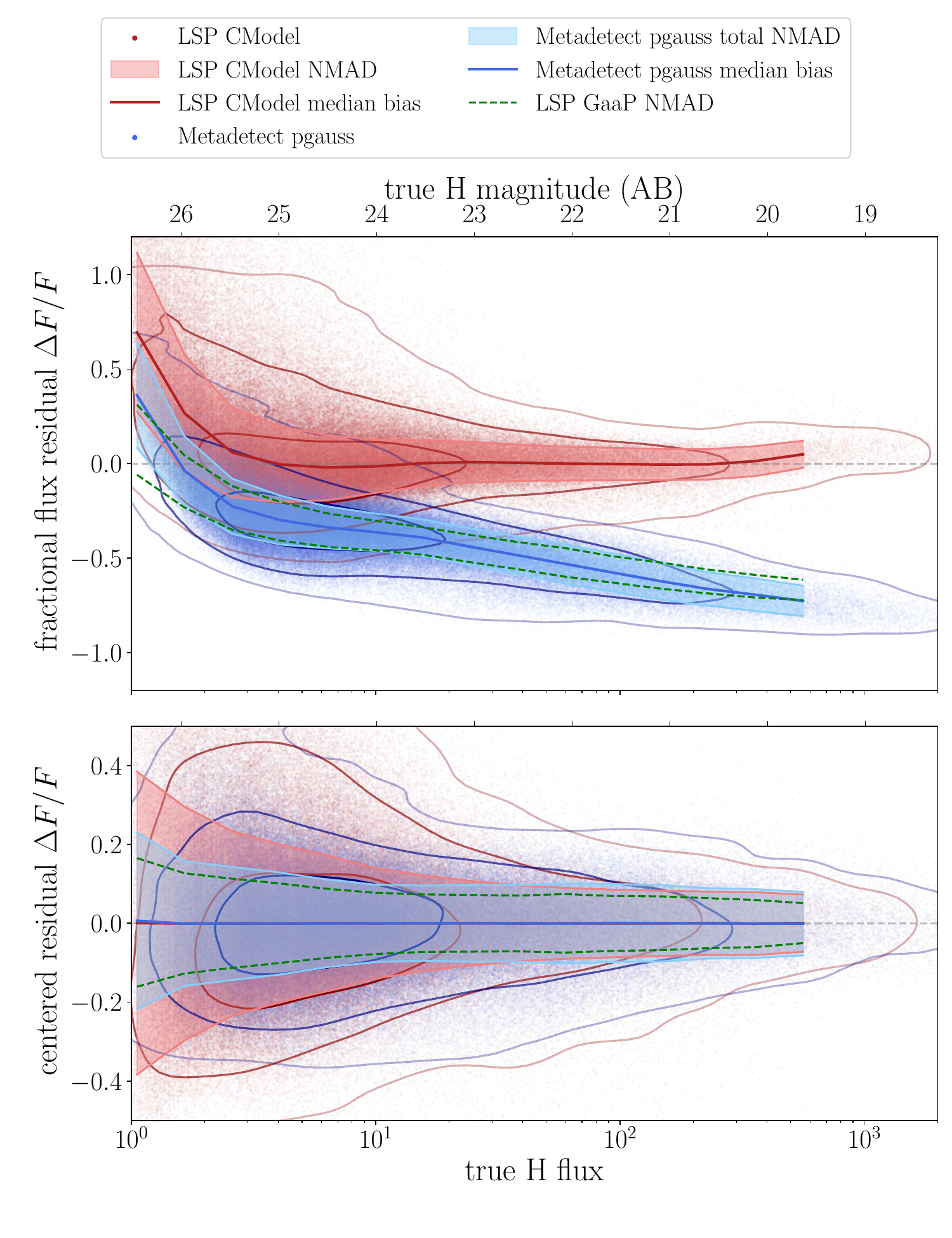}
    \caption{\textit{top}: The fractional residuals of the measured fluxes in the H band using the LSST Science Pipelines CModel algorithm (red), and the \texttt{Metadetection} \texttt{pgauss} algorithm (blue). The shades in the corresponding colors represent the NMAD scatter as a function of true flux. We also show the NMAD scatter from the GAaP photometry from the LSST Science Pipelines with green dashed lines. \textit{bottom}: Similar to the top panel, but with the median residual subtracted. The contours in the corresponding colors represent the 1, 2, and 3$\sigma$ ranges in 2D. Note that while the \texttt{Metadetection} photometry (blue) shows a clear bias as expected (upper panel), its scatter reaches a similar systematic floor to that of LSP photometry on the brighter end (bottom panel). The \texttt{Metadetection} photometry, on the other hand, exhibits a smaller level of scatter on the fainter end. Also note that the GAaP photometry (green) shows a similar behavior to that of \texttt{Metadetection} \texttt{pgauss}, while the scatter is smaller.}
    \label{fig:flux_residual_compare}
\end{figure}

In Fig.~\ref{fig:flux_residual_lsp}, we show the fractional flux residual (measured flux / true flux - 1) as a function of true flux (lower axes) and true AB magnitude (upper axes), in the three Roman bands as indicated in the upper right corners of each panel. The red points correspond to individual objects, and the blue bands represent the 1$\sigma$-equivalent ranges using normalized median absolute deviation (NMAD = 1.4826 $\times$ median absolute deviation). Overall, the measured flux is consistent with the true flux at all magnitudes, except for the expected selection effect primarily in the reference band (H) below the 5$\sigma$ limiting magnitude \citepalias[in this case $\sim$25.3 in H for extended objects;][]{Rotac}, resulting in an upturn of the flux bias towards the dimmer end. The measured NMAD scatter at a magnitude of 25 is $\sim$20\%, and $\sim$8\% at a magnitude of 20. Note that the flux scatter asymptotes to the $7-8$\% level at the brighter end with a slight upturn, implying a systematic uncertainty floor mainly due to modeling biases. 

Also plotted are the root mean square (RMS) errors of the fractional flux residual: errors with the noise correlation accounted for in cyan and diagonal errors in navy. As mentioned in Sec.~\ref{subsec:correlated_noise}, not accounting for the noise correlation in the flux uncertainty results in $\sim$4 times smaller flux uncertainty, overestimating the measured SNR. 

As mentioned in Sec.~\ref{subsec:photometry_mdet}, due to the current limitations of our software, we do not account for the noise correlations between pixels when calculating the flux errors for the \texttt{pgauss} photometry. Therefore, the flux SNRs in our \texttt{Metadetection} catalog are overestimated by a factor of few (Sec.~\ref{subsec:correlated_noise}). To compensate for the effect of this on our SNR $>5$ selection, we apply a different SNR cut on the \texttt{Metadetection} catalog, as follows. With the LSP selection (near the beginning of this section), the final LSP catalog reaches 50\% completeness (compared to the truth population) at a true AB magnitude of $\sim$25.9 in H. At this depth, the well-matched \texttt{Metadetection} galaxies have a median SNR of 17.1. We therefore adopt a \texttt{pgauss} SNR $>17.1$ threshold, in addition to the other filters that are similar as those of the LSP, so that its selection turnover approximately matches the 50\% completeness depth of the LSP CModel SNR $>5$ sample. The final number of the selected \texttt{Metadetection} sample is 278,017 compared with 273,451 for the LSP catalog (less than 2\% difference in number of sources), showing the validity of our approximation.

In Fig.~\ref{fig:flux_residual_compare}, we compare the photometry from the LSP (red) and \texttt{metadetection pgauss} (blue) in the H band. Note that the other bands exhibit a similar behavior. On the top panel, we plot again the fractional flux residual. As opposed to the LSP photometry, the \texttt{pgauss} result shows gradually increasing flux underestimation towards the brighter end, by more than 50\% at magnitude $<22$. However, this is a known effect \citep{Yamamoto2025}, as achieving unbiased photometry is not the main goal of \texttt{metadetection} photometry. 

The bottom panel shows the ``median-subtracted'' fractional flux residuals. On the dimmer end, the \texttt{pgauss} photometry shows somewhat smaller scatter than that of the LSP: a $\sim$35\% smaller NMAD scatter at an H-band magnitude of 25. However, more interestingly, the scatter from the two algorithms asymptotes to a similar value on the brighter end (7-8\%), implying that neither of them is dominated by pure statistical uncertainty and therefore it may be worth exploring whether there is a similar source for this shared systematic floor. 

\begin{figure}
    \includegraphics[width=\columnwidth]{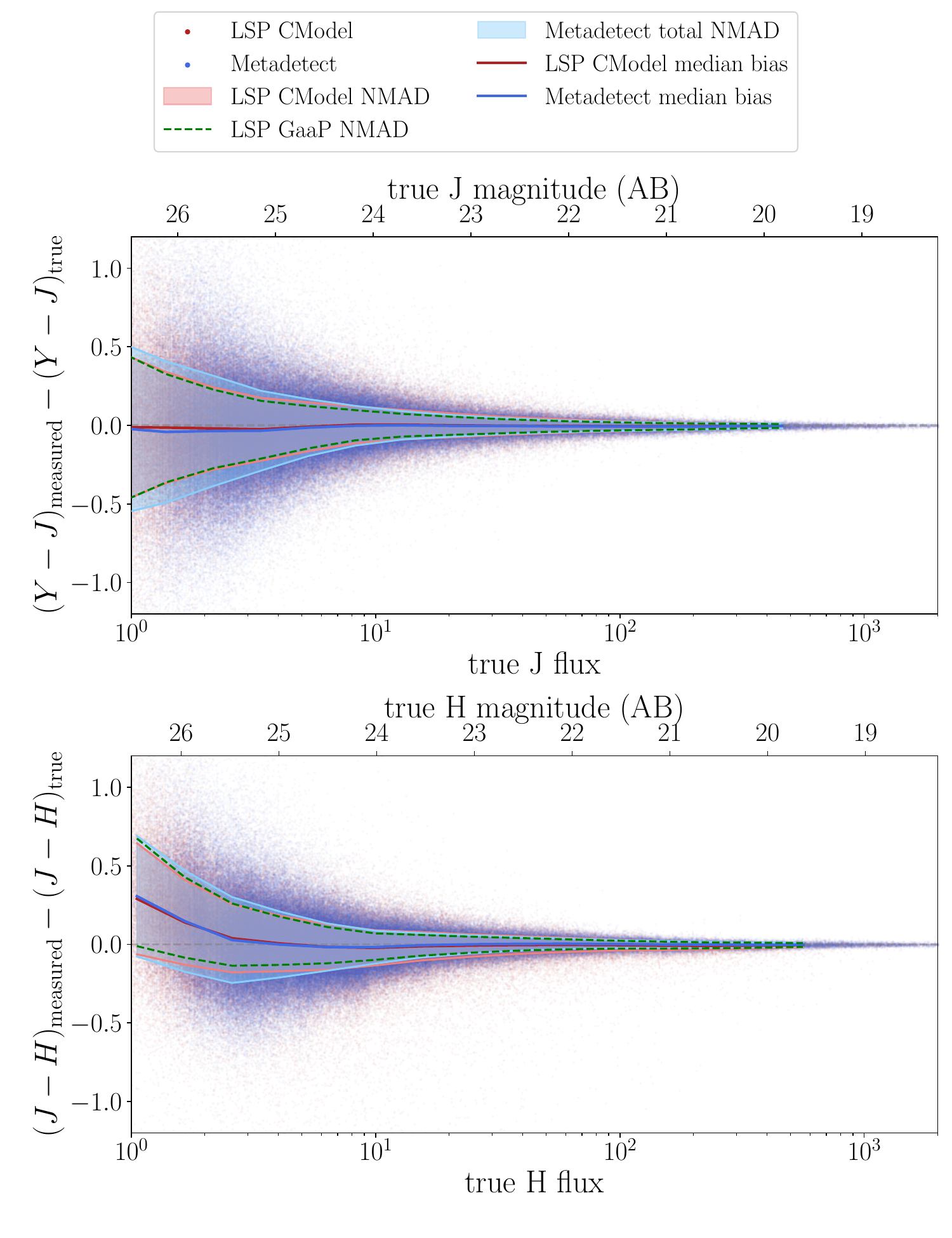}
    \caption{The color residuals of $Y-J$ (top), and $J-H$ (bottom) and their NMAD (shades) as a function of true flux. The red colors correspond to the LSST Science Pipelines (LSP) CModel algorithm and the blue colors to the \texttt{Metadetection} \texttt{pgauss} algorithm. We also show the results from the GAaP photometry from the LSP in green. Note that although the \texttt{pgauss} photometry shows a clear flux bias as shown in Fig.~\ref{fig:flux_residual_compare}, it shows almost indistinguishable performance in color retrieval compared with that of LSP CModel. Also note that while the flux scatter in Fig.~\ref{fig:flux_residual_compare} reaches a systematic uncertainty floor at the brighter end, there is no apparent systematic floor for the colors plotted here. Lastly, the scatter in color residuals with the LSP GAaP photometry shows a similar level of performance to that of the LSP CModel, despite its smaller scatter in flux residuals (Fig.~\ref{fig:flux_residual_compare}, bottom).}
    \label{fig:color_residual_compare}
\end{figure}

When estimating photo-$z$s of galaxies, more fundamental quantities than the galaxy magnitudes are the colors of the galaxies. In Fig.~\ref{fig:color_residual_compare}, we compare the distribution of the color residuals (measured color - true color) between \texttt{Metadetection pgauss} and the LSP CModel photometry. On the top panel, we plot the result for the Y-J color, and on the bottom panel, that of the J-H color. While both algorithms recover the true colors on average, a noticeable observation is that in spite of the significantly different levels of flux residuals between the two algorithms especially towards the dimmer end (Fig.~\ref{fig:flux_residual_compare}, bottom panel), the color residuals do not exhibit any significant difference (even slightly larger for \texttt{pgauss}). It implies that the correlation between the flux biases across different bands is weaker for \texttt{pgauss} photometry than the LSP CModel photometry. Given that the galaxy colors are the most important factor for the photo-$z$ estimation, one expects similar performance of the photo-$z$ estimation between the two photometry algorithms. Another important observation is that unlike the flux measurement, the scatter in the color residuals asymptotes to a negligible level towards the brighter end. It strongly suggests that the aforementioned 7-8\% systematic flux errors on the brighter end are highly correlated between different bands, i.e., a systematic flux bias independent of wavelength. This could occur, for example, due to a model bias arising from a mismatch between the photometry model and the real surface brightness profile, since the same model is used for all bands.

In Fig.~\ref{fig:flux_residual_compare} and \ref{fig:color_residual_compare}, we plot the results with the LSP GAaP photometry in green, for additional comparison. For the flux residual, the GAaP photometry shows a similar level of flux bias to the \texttt{Metadetect pgauss} photometry (the top panel of Fig.~\ref{fig:flux_residual_compare}), while showing a somewhat smaller NMAD scatter (e.g., $\sim$22\% smaller at a H-band magnitude of 25, bottom) than that of \texttt{pgauss}. This is an expected behavior because both measure PSF-corrected, Gaussian-weighted aperture fluxes designed to provide consistent effective apertures across bands. On the other hand, the GAaP colors (Fig.~\ref{fig:color_residual_compare}) show a very similar level of residuals as that of the LSP photometry, with a slightly smaller scatter for J$-$H colors towards the fainter end. Overall, this comparison confirms the similarity between the \texttt{pgauss} and GAaP photometry algorithms and the validity of our choice to use GAaP photometry for color estimation.

\begin{figure}
    \includegraphics[width=\columnwidth]{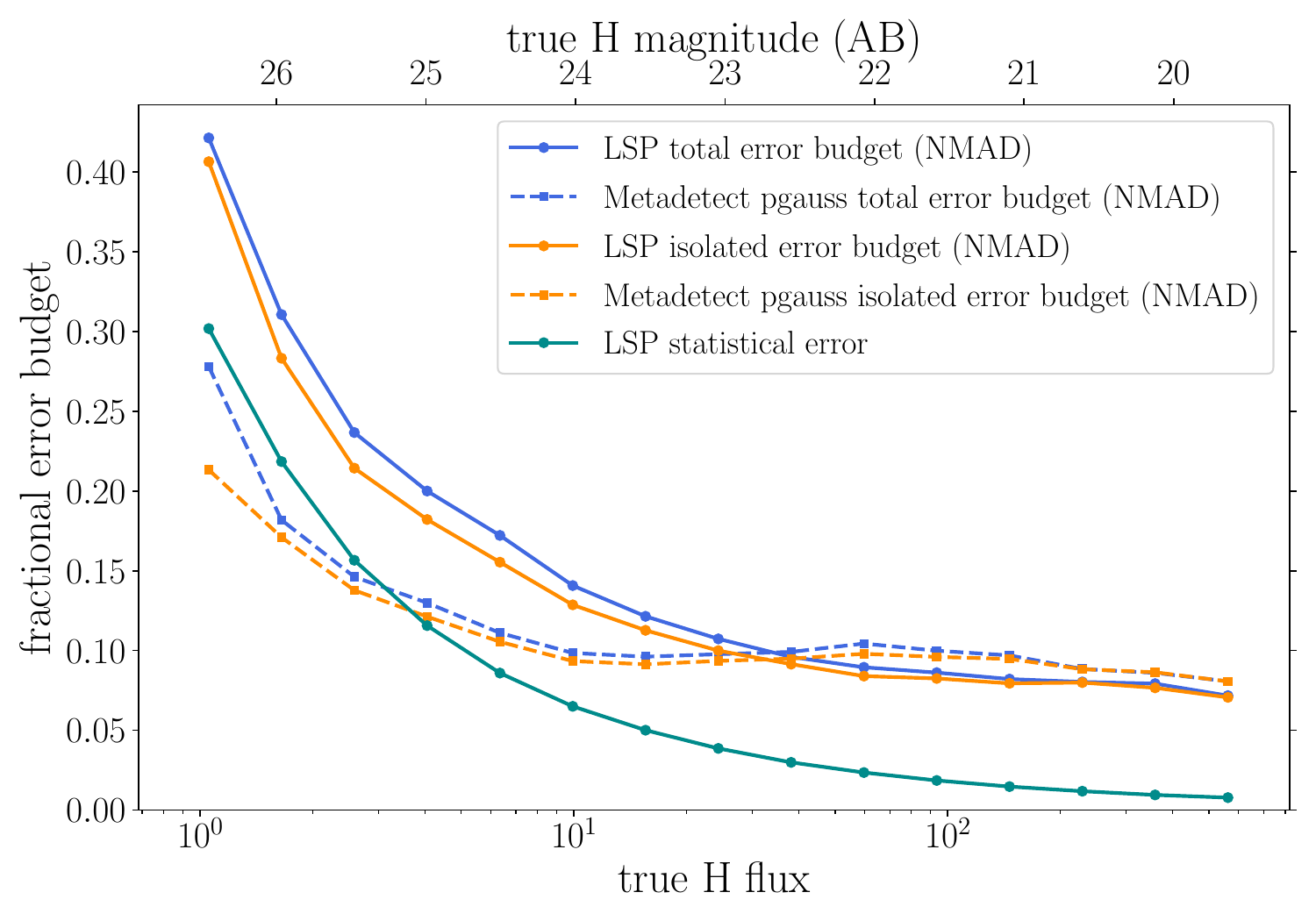}
    \caption{The fractional flux error budget for total empirical scatter (blue, NMAD), for isolated galaxies (orange, NMAD), and for the statistical scatter (green, RMS). The solid lines correspond to the CModel photometry from the LSST Science Pipelines, and the dashed lines to the \texttt{pgauss} photometry from \texttt{Metadetection}. Note that 1) the statistical error (green) explains a significant fraction of the total error budget (blue) on the dimmer end, but is negligible on the brighter end, 2) blending does not account for a significant fraction of the total error budget (blue vs. orange), and 3) the two photometry methods, in spite of their wildly different flux measurement algorithms, asymptote to a similar level of systematic error ($\sim$7-8\%) floor on the brighter end.}
    \label{fig:error_budget}
\end{figure}

Next, we study the contribution of the different sources of errors in the magnitudes. We focus on two comparisons: the statistical error vs.\ the total error, and the error from the isolated objects vs.\ the total error. We define isolated galaxies as those that have no other true object within 2$\arcsec$ brighter than 5\% of the H flux of the target galaxies (magnitude difference of $\sim$3.25 magnitudes). We have tested that increasing this 2$\arcsec$ aperture to a reasonable number does not meaningfully change our result. As a result, we select 134,340 isolated galaxies out of the total 273,451 (49\%). In Fig.~\ref{fig:error_budget}, we plot the 1$\sigma$ fractional flux errors as a function of true flux and magnitude in the H band. In solid lines are the results from the LSP CModel photometry. The green line shows the RMS error of the statistical uncertainties of the objects (the correlated noise, Eq.~\ref{eq:corr_error}). At low fluxes, the statistical error occupies a significant fraction of the total error (the blue solid line, NMAD, corresponding to the blue shade in Fig.~\ref{fig:flux_residual_lsp}). However, at larger fluxes, the total error budget becomes dominated by systematics, with the statistical error being negligible with respect to the total, which could also be seen in Fig.~\ref{fig:flux_residual_lsp}.

The orange line corresponds to the NMAD scatter of the isolated galaxies. Note that the scatter of the isolated galaxies explains more than 90\% of the total scatter at all magnitudes. This finding suggests that the blending of the objects is not a dominant source of the systematic flux error, and that it is an attribute of all galaxies (such as mismatches between the true and fit profiles) that is responsible for the systematics floor. 

Meanwhile, the \texttt{Metadetect pgauss} photometry (dashed lines in the corresponding colors) exhibits behavior similar to that of the LSP CModel. The fact that the two distinct algorithms show a similar level of systematic uncertainties at the brighter end hints at a similar source of systematic errors such as mismatched modeling. 

\begin{figure}
    \includegraphics[width=\columnwidth]{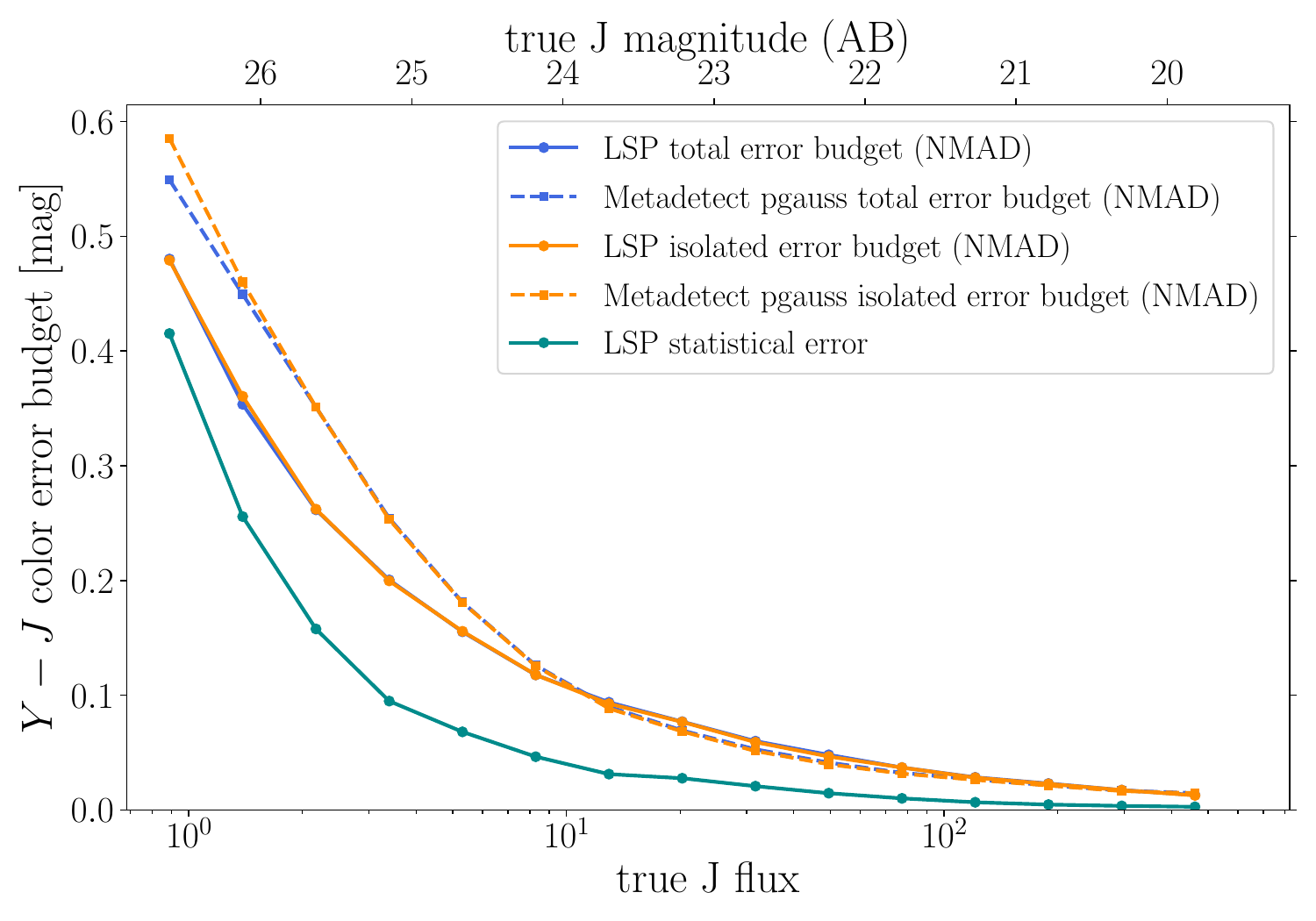}
    \caption{Similar to Fig.~\ref{fig:error_budget}, but for Y$-$J color. Note that 1) the statistical scatter asymptotes to nearly zero at the brighter end, as opposed to the flux error case, and 2) for the dimmest magnitude bin, the statistical uncertainty explains $\sim$75\% of the total variance (compared with $\sim$51\% for the flux error case).
    }
    \label{fig:error_budget_color}
\end{figure}

In Fig.~\ref{fig:error_budget_color}, we show the corresponding 1$\sigma$ errors in the Y$-$J color as a function of the true J-band flux (lower ticks) and magnitude (upper ticks). As mentioned, the total scatter asymptotes to a small value close to zero towards the brighter end (the remaining systematic budget being only $\sim$5\% of the median color, 0.28, in the brightest bin). Also, in the dimmest bin, the statistical uncertainty (green) explains approximately 75\% of the total variance, compared with 51\% for the flux residual case (Fig.~\ref{fig:error_budget}). Interestingly, note that galaxy blending does not show any discerning impact on the color uncertainty (yellow vs. blue; c.f., for the flux residuals, there is a distinguishable level of blending effect). Moreover, we confirm our previous finding that the \texttt{pgauss} photometry (dashed) and the LSP photometry (CModel, solid) merge to almost the same level of color scatter at the brighter end.

Together with our finding that the LSP and \texttt{Metadetect pgauss} photometry algorithms show a similar level of performance in measuring galaxy colors given the specifications of our Roman images, we focus on the LSP photometry from now on.

\subsection{Results for Rubin photometry}
\label{subsec:photometry_result_rubin}

As described in Sec.~\ref{subsec:photometry_lsp}, we perform forced photometry in all six bands on Rubin coadded images that overlap with our 1.23 deg$^2$ of simulated Roman footprint. Therefore, the selection is identical to that from the previous section. However, for the Rubin color analysis for this section, we exclude those that have negative or \texttt{NaN} fluxes.\footnote{Note that for the subsequent photometry error modeling (Sec.~\ref{sec:error_modeling}), we include negative fluxes, only excluding NaN values, which affect at most 1,222 galaxies out of our $\sim$270k galaxies (in the \textit{u} band).}  

\begin{figure*}
    \includegraphics[width=\textwidth]{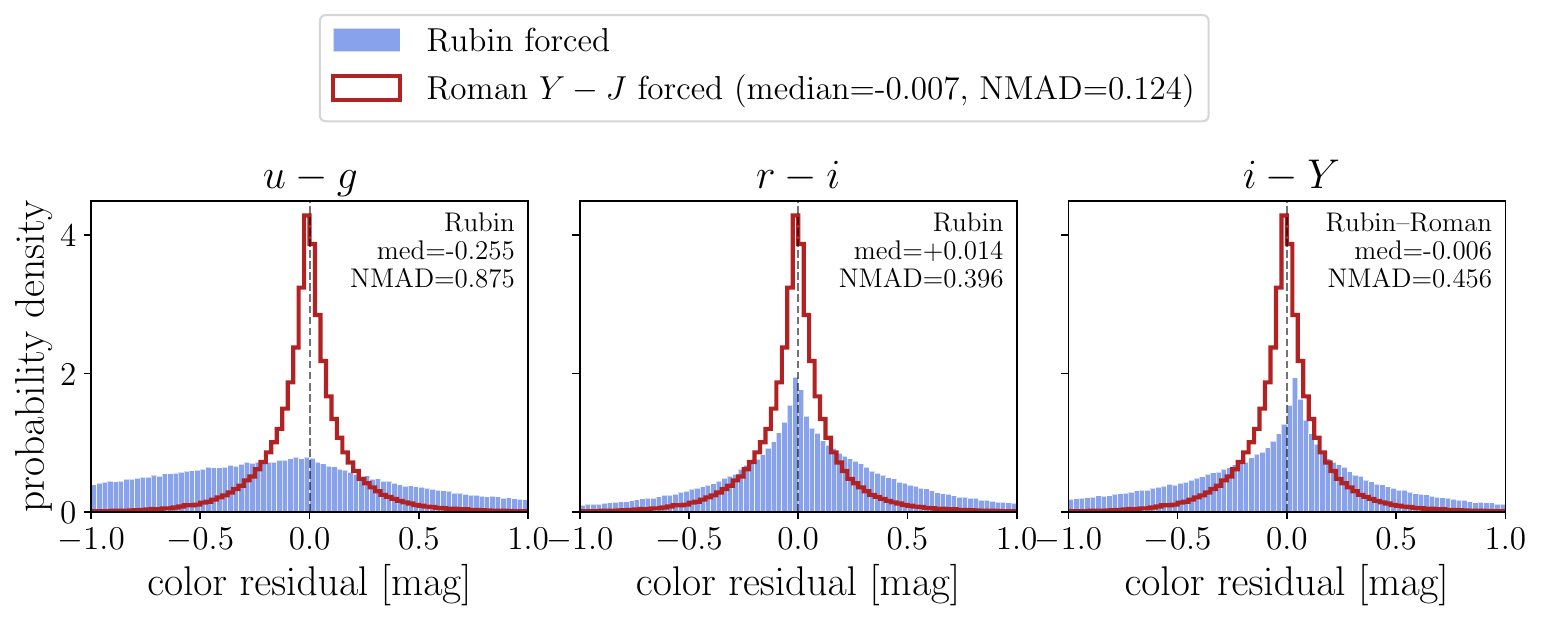}
    \caption{The histograms of color residuals with respect to the truth for two forced Rubin colors ($u-g$, $r-i$) and one Rubin--Roman color ($i - {\rm Y}$) using GAaP photometry (blue, Sec.~\ref{subsec:photometry_lsp}). The red histogram represents the color residual distribution for the forced Roman $Y-J$ color (also GAaP), for comparison. For each color, we also report the median bias and the NMAD scatter. Note that we only show the two Rubin colors, since the residuals of the other three Rubin colors qualitatively follow the trend in \textit{r-i} colors.}
    \label{fig:color_hist_rubin}
\end{figure*}

In Fig.~\ref{fig:color_hist_rubin}, we show the color residuals for the Rubin forced photometry (Sec.~\ref{subsec:prepare_rubin_images}) with respect to the truth in two Rubin band combinations, \textit{u$-$g} (left) and \textit{r$-$i} (middle), and one Rubin-Roman combination, $i$$-$Y (right). We also plot the Y$-$J color residual for the Roman forced photometry in red for comparison. Note that the other three Rubin colors qualitatively follow the trend shown in \textit{r$-$i} color. The measured color residual distribution in \textit{u$-$g} color shows a noticeably heavier tail towards blue (negative residuals). This is likely due to the combination of 1) the large distance of \textit{u} band from the reference H band, which results in a significant fraction of negative fluxes that are excluded from the plot ($\sim$23\%; compared with $\sim$5\% in \textit{i} band) therefore biasing the \textit{u} band magnitudes low on average, and 2) the correlation of the flux errors relative to the truth between adjacent Rubin bands being the weakest for \{\textit{u, g}\} combination (correlation coefficient of $\sim$0.7, compared with $>0.9$ for other adjacent colors), which isolates the effect 1) in the \textit{u} band. As a result, the \textit{u$-$g} color shows a significantly larger NMAD scatter (0.875) than other colors (typically 0.4-0.6).\footnote{However, note that the \textit{u-g} color residual distribution still peaks at zero.} On the other hand, the \textit{r$-$i} color residuals show a much more symmetric distribution about zero, with an NMAD scatter of 0.396, approximately 3 times larger than that of the Y$-$J color residuals in red which we show for comparison. Overall, we confirm that we achieve unbiased color distributions on average in Rubin bands with respect to the truth except for the aforementioned caveat in the \textit{u$-$g} color. We use these force-measured fluxes in Rubin coadded images in combination with those in the three Roman bands in the subsequent analysis. 

On the other hand, the $i$$-$Y color shows a peak that positively deviates from zero by $\sim$0.05 magnitude. This is an expected behavior where the substantially different image resolutions and PSFs of Rubin and Roman prevent the nominally matched GAaP apertures from sampling identical regions of each galaxy, thereby breaking the method’s assumption of consistent aperture photometry across bands. Note that this is not likely to cause any significant bias in photo-$z$ estimation for algorithms that use machine-learning methods, as long as one has a representative training/spectroscopic sample as in this analysis, while it could non-negligibly affect the results from template-fitting-based photo-$z$ estimation methods.

\section{Modeling of photometry error}
\label{sec:error_modeling}

As briefly mentioned in Sec.~\ref{subsec:OU2024}, the \texttt{Diffsky} mock galaxy catalog used for our simulated Rubin and Roman images is subject to a caveat: the repeated SEDs and the lack of variance in NIR colors results in an unrealistically high flux correlation between different bands and therefore overly-optimistic photo-$z$ estimates \citep[][also see Fig.~2 in \citealt{Berlfein2026}]{OU2024}.

To mitigate this limitation, using a set of machine-learning (ML) techniques, we model the measured photometry errors as a function of true galaxy properties in the OpenUniverse2024 simulations -- and then we apply this model to a newer version of the \texttt{Diffsky} mock catalog with improved realism in the galaxy colors (Sec.~\ref{subsec:diffsky}). This approach enables us to carry out catalog-level studies of photometric redshift estimation using photometry with greater realism in the SEDs, and with realistic systematic and statistical uncertainties. In this section, we describe our training algorithm, the corresponding results, and the theoretical error calculation for comparison. Note that we select galaxies in the new \texttt{Diffsky} mock catalog with redshifts between 0 and 3 for this analysis. Given one set of mock photometry with the known truth, our algorithm provides a convenient way to reproduce realistic object flux errors without having to run additional expensive image simulations, processing, detection, and measurement steps including photometry. Also note that our algorithm is limited to a given observation depth and detector configuration.

\subsection{Training algorithm}
\label{subsec:error_training}

We start by collecting the measured fluxes and the corresponding true fluxes in all nine Rubin and Roman bands, using GAaP photometry (for galaxy colors, as mentioned earlier) with the sample selection described in Sec.~\ref{subsec:photometry_result_roman}. We exclude a small fraction ($<$0.1\%) of fluxes that are NaN in at least one band, which does not meaningfully affect the result. 

Our target variables are fractional flux residuals in each band, $b$:
\begin{equation}
\label{eq:Y_data}
    Y_b \equiv \frac{F_{{\rm measured}, b}}{F_{{\rm true}, b}} - 1.
\end{equation} 
We use a set of histogram-based gradient-boosted decision trees (HGBDT) as implemented in \texttt{scikit-learn}\footnote{https://scikit-learn.org/} \citep{scikit-learn} to train the fractional flux errors as a function of true galaxy properties (magnitude in each band, H-band disk size, H-band bulge size, and H-band disk fraction: $X$). A HGBDT discretizes continuous input features into a finite number of bins, allowing candidate node splits within each decision tree to be evaluated using the summed derivatives of the loss function for the samples in each bin, rather than by testing every input feature value. As in conventional gradient boosting, decision trees are added sequentially to reduce the residual prediction errors of the current ensemble, while the histogram approximation substantially lowers computational and memory costs with minimal loss of predictive accuracy. 

We randomly split the catalog into 80\% for training and 20\% for testing. The model is trained in three stages. First, for each band, we train an HGBDT model to predict the conditional median flux bias, $\hat{B}_b(X)$, and define the remaining residual as $R_b = Y_b - \hat{B}_b(X)$. Second, we train separate HGBDT quantile regressors at 23 quantiles (ranging from 0.005 to 0.995) to model the conditional residual distribution, $\hat{Q}_{R_b}(u|X)$, with  a linear interpolation (u: percentile value). Finally, to train the correlation of flux residuals between bands, we transform the estimated percentile ranks of the training residuals into a Student-$t$ latent space with the degrees of freedom of 4 and estimate magnitude-dependent local correlation matrices using the 500 nearest training galaxies in standardized magnitude space. We model correlations separately among the six Rubin bands and among the three Roman bands, while neglecting Rubin--Roman correlations because we find them to be insignificant.

We apply the trained model to the testing sample and the new \texttt{Diffsky} mock catalog by first predicting $\hat{B}_b(X)$ and the 23 conditional residual quantiles for each galaxy. We then use its true magnitudes to determine the corresponding local correlation matrix, draw a correlated multivariate Student-$t$ realization, and transform each component into a percentile of the predicted residual distribution. Piecewise-linear interpolation between the predicted quantiles gives a residual realization, $R_{b,\mathrm{draw}}$, to which we add the predicted median bias to obtain the final fractional flux error realization, $Y_{b,\mathrm{draw}}=R_{b,\mathrm{draw}}+\hat{B}_b(X)$. The emulated measured flux is then calculated as $F_{\mathrm{measured},b}=F_{\mathrm{true},b}(1+Y_{b,\mathrm{draw}})$, thereby reproducing the property-dependent flux bias, residual scatter, and correlations between bands without rerunning the image simulations and photometry pipelines.

In Appendix~\ref{app:diagram}, we provide further implementation details and show a schematic diagram that summarizes these training and estimating procedures.

\subsection{Results}
\label{subsec:error_modeling_result}

\begin{figure*}
    \includegraphics[width=\textwidth]{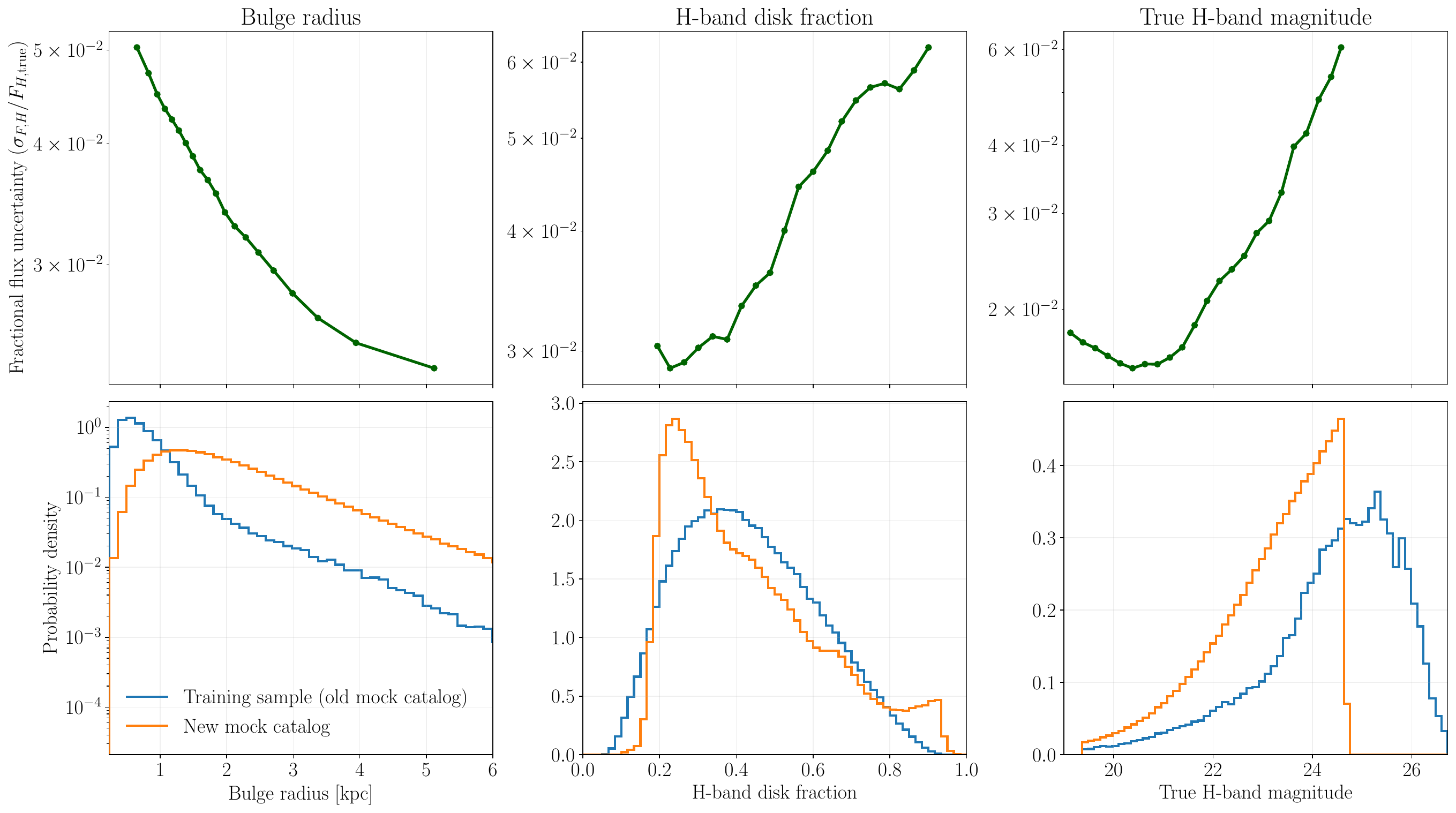}
    \caption{\textit{Top}: The trained fractional flux uncertainty (median) in the H band as a function of input truth variables (Sec.~\ref{subsec:error_training}): bulge half-light radius (left), H-band disk fraction (middle), and true H-band magnitude (right). Note that the flux uncertainty decreases with increasing galaxy sizes (bulge half-light radius; disk half-light radius exhibits a similar trend) as expected (brighter objects). Also, it shows a positive correlation with disk fraction (a higher disk fraction generally corresponding to a more extended object given a flux) and the H-band magnitude (as expected). \\
    \textit{Bottom}: The corresponding distributions of each galaxy property for the training sample (old \texttt{Diffsky} mock catalog; blue) and the new mock catalog we use for the photo-$z$ analyses (orange). Note that the property space of the training sample covers that of the new mock catalog to which we apply our model.
    }
    \label{fig:empirical_error_vs_properties}
\end{figure*}

In Fig.~\ref{fig:empirical_error_vs_properties}, we show the dependence of the modeled fractional flux uncertainty in the H band, $\sigma_{F,{\rm H}}/F_{\rm H,true}$ on the input truth variables. In the top left panel, we plot the flux uncertainty vs.\ bulge half-light radius in kpc. Not surprisingly, the fractional flux uncertainty decreases for larger (therefore more likely to be brighter) objects; disk half-light radius exhibits a similar trend. Meanwhile, it shows a positive correlation with the H-band disk fraction (top middle panel). This is because, as the disk fraction increases, the galaxy size also increases, covering a larger area (more pixel noise included) for a given total flux. Lastly, while the fractional flux uncertainty is overall an increasing function of the true H-band magnitude (top right panel) as expected, it shows an upturn around the magnitude of $\sim$20, where the uncertainty is dominated by systematics and the sample is sparse (Sec.~\ref{subsec:photometry_result_roman}). In the bottom panels, we show the distributions of the corresponding galaxy properties for our training sample (old \texttt{Diffsky} mock catalog) in blue and the new mock catalog to which we apply our model, in orange. Note that the property space of the training sample covers that of the new mock catalog.

\begin{figure*}
    \includegraphics[width=\textwidth]{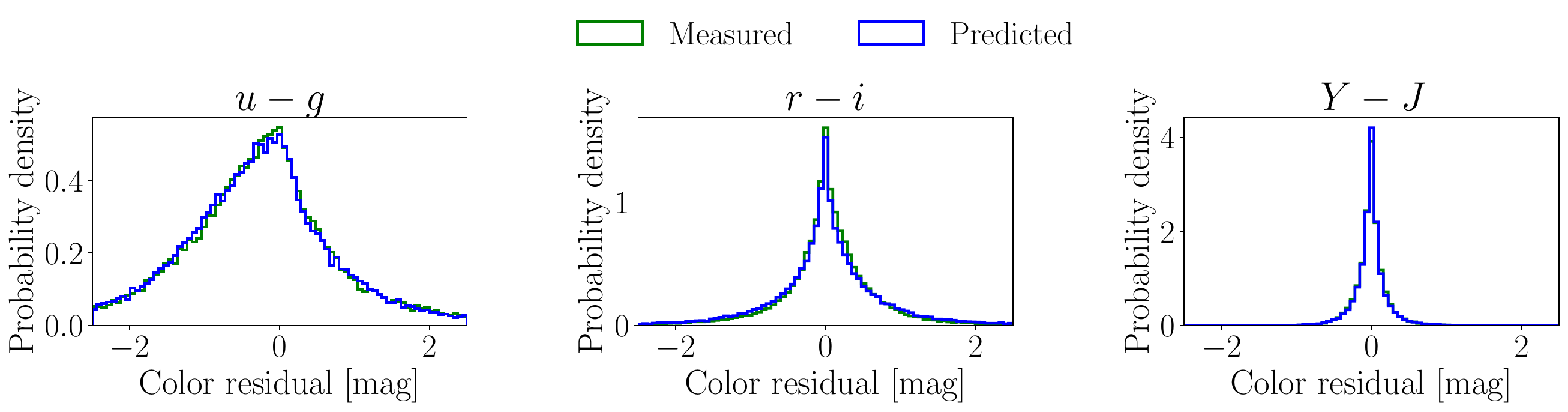}
    \caption{The distributions for the measured (Sec.~\ref{sec:photometry}, green) and model-predicted (Sec.~\ref{subsec:error_training}, blue) colors: two Rubin colors ($u-g$, $r-i$), and one Roman color ($Y-J$) as a representative sample. The predicted color distributions reasonably follow that of the measured. Note the skewness in the $u-g$ colors, for which a possible reason is given in Sec.~\ref{subsec:photometry_result_rubin}.}
    \label{fig:color_distributions_model_data}
\end{figure*}

In Fig.~\ref{fig:color_distributions_model_data}, we plot the result from the error modeling in the previous section using our testing sample. Since our main interest is the prediction performance of the color errors, we plot three representative colors: $u-g$ and $r-i$ colors for Rubin, and $Y-J$ color for Roman. Note that the other Rubin colors closely follow the trend shown in the $r-i$ color plot, and similarly for the Roman colors. The blue histograms represent the color residuals with respect to the true colors predicted by our model, and the actual measured color residuals from the testing sample are plotted in green.

First, our ML model accurately predicts the highly skewed distribution in the $u-g$ color (the left panel; as in Sec.~\ref{subsec:photometry_result_rubin}). Other Rubin colors, representatively shown with $r-i$ color in the middle panel, also closely follow the distributions of the measured colors of the testing sample. Using the objects between the 2.5th and 97.5th percentile in color residual (to remove outliers), the median and the standard deviation of the measured distribution are -0.26 (0.02) and 0.95 (0.51) for the $u-g$ ($r-i$) color, respectively, compared with -0.29 (0.00) and 0.96 (0.55) for the predicted distribution. Furthermore, the Roman colors, represented by $Y-J$ color on the right panel, also show excellent agreement between the model-predicted and measured true colors (median and standard deviation being almost identical). This agreement between our model prediction and the true measured colors confirms the validity of our ML modeling of the photometry errors in each band as well as that of the correlations between different bands.\footnote{Since we need an accurate correlation between two bands to retrieve the unbiased color distribution given that it is the difference in magnitude between two bands.}

\begin{figure}
    \includegraphics[width=\columnwidth]{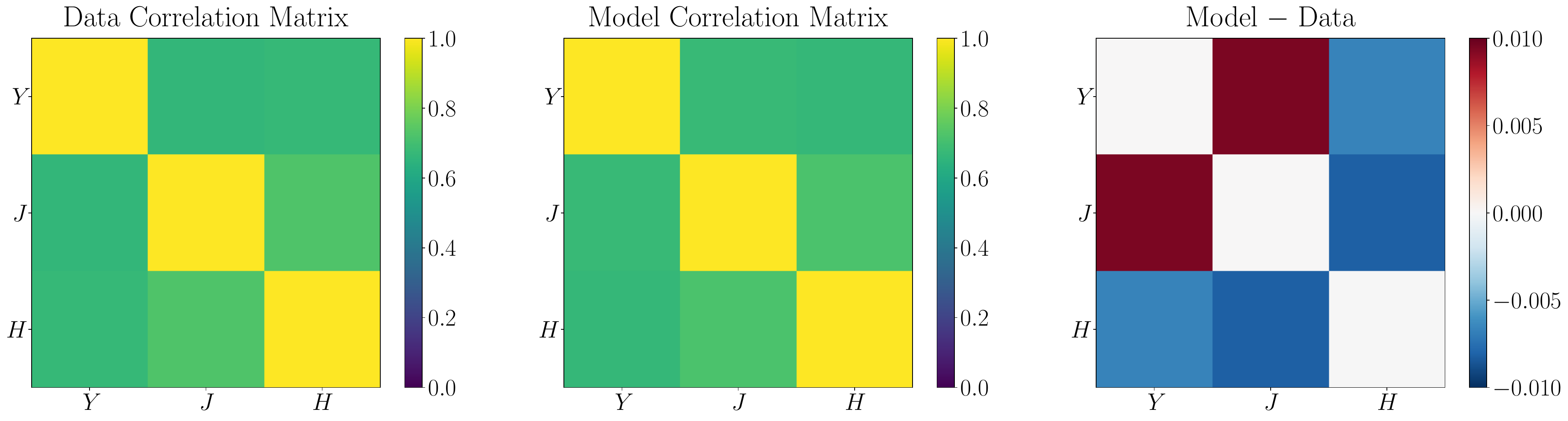}
    \includegraphics[width=\columnwidth]{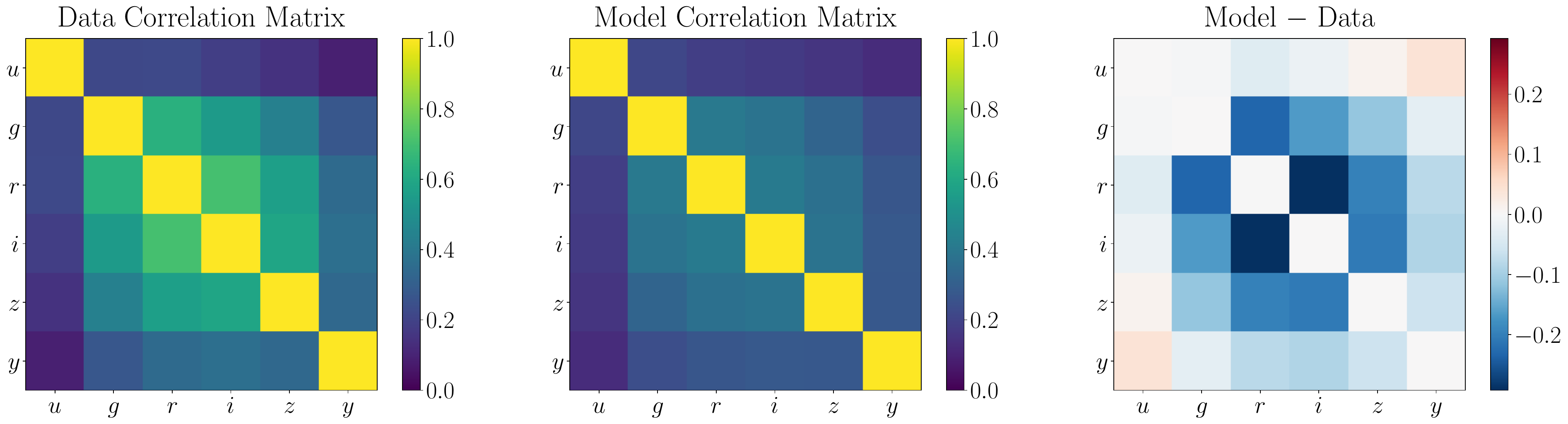}
    \caption{The correlation matrices between the fractional flux residuals of different Roman (top) and Rubin (bottom) bands. The correlations from the measured objects (Sec.~\ref{sec:photometry}) are shown in the left panel, and those from the model-prediction (Sec.~\ref{subsec:error_training}) are shown in the middle panel. Shown in the right panels are the differences between the model predictions and the measured values. Note that for the Roman case, the correlation coefficients from the model prediction closely follow those from the measured data (correlation coefficients of $\sim$0.7-0.8, agreeing within 0.01), while there is moderate underestimation of the correlation coefficients for the Rubin bands (predicted correlations of 0.2-0.4, compared with truth of 0.2-0.7; up to the difference of $\sim$0.3). However, the reasonably similar color distributions between the model predictions and the measured data (Fig.~\ref{fig:color_distributions_model_data}) suggest that the underestimation of the correlation coefficients is likely due to a small fraction of color outliers or slight data skewness that the model cannot properly capture by construction.}
    \label{fig:error_correlation}
\end{figure}

In Fig.~\ref{fig:error_correlation}, we plot the correlation matrices (normalized covariance matrix) of the model-predicted fractional flux errors between different bands (Eq.~\ref{eq:Y_model}, middle panels) and those of the testing sample (Eq.~\ref{eq:Y_data}, left panels). The Roman results are plotted in the top panels and the Rubin results are plotted in the bottom panels. When calculating the correlation coefficients, we limit the fractional flux errors between the 2.5th and 97.5th percentile of the testing sample range, by which we remove outliers to robustly measure the correlation within a $\sim$$2\sigma$ range, which dominates our results.

For Roman bands, the model is very effective at predicting the correlation of the flux errors (the top two panels), exhibiting correlations of $\sim$0.7-0.8 between bands. The differences in the correlation coefficients are less than 0.01 in all bands (the top right panel). On the other hand, our model somewhat underestimates the flux error correlations especially in $g, r, i, z$ bands (by up to $\sim$0.3 in correlation coefficient; 0.2--0.7 truth compared with 0.2--0.4 prediction; the bottom right panel). However, given the good agreement (between measured vs.\ predicted) of the color residual distributions shown in Fig.~\ref{fig:color_distributions_model_data}, which involves the correlation between the two used bands, this moderate discrepancy is not likely to affect our final photo-$z$ results significantly. Also, it implies that this discrepancy may originate from outliers or slight data skewness (see the subtle difference between the two distributions in the middle panel of Fig.~\ref{fig:color_distributions_model_data}) that our model cannot fully capture.

\subsection{Calculation of theoretical flux errors}
\label{subsec:error_theoretical}

In addition to the empirical error model trained in the previous section, we calculate the expected theoretical (statistical) flux errors to compare the expected photo-$z$ outcomes 1) between different LSST completed years, and 2) between the empirical error model and the theoretically calculated errors. For the Roman bands, we start from the point-source 5$\sigma$ limiting magnitudes for the HLWAS medium tier reported by \citetalias{Rotac} converted to the point-source flux noise using the relation
\begin{equation}
    \frac{F_5}{\sqrt{F_5 + V_{\rm point,PSF}}} = 5,
\end{equation}
where $F_5$ is the 5$\sigma$ limiting flux, which becomes the self-Poisson flux noise in the denominator.
For the Rubin bands, we start from the per-pixel flux variance, $V_{\rm pix}$, calculated in Sec.~\ref{subsec:prepare_rubin_images}. This per-pixel total background noise variance for Rubin bands, $V_{\rm pix}$, is converted to the PSF-size-corrected point-source count variance ($V_{\rm point,PSF}$) as $V_{\rm point,PSF} = N_{\rm eff} \times V_{\rm pix}$, where the effective number of pixels $N_{\rm eff}$ is $2.266 \times {\rm FWHM}_{\rm eff}^2$ which comes from the Gaussian noise-equivalent area \citep{Jones2021}. As in Sec.~\ref{subsec:prepare_rubin_images}, we take the ${\rm FWHM}_{\rm eff}$ values for each HEALPix pixel with $N_{\rm side}=64$ and take the corresponding median value for our fiducial choice. 

\begin{figure*}
    \includegraphics[width=\textwidth]{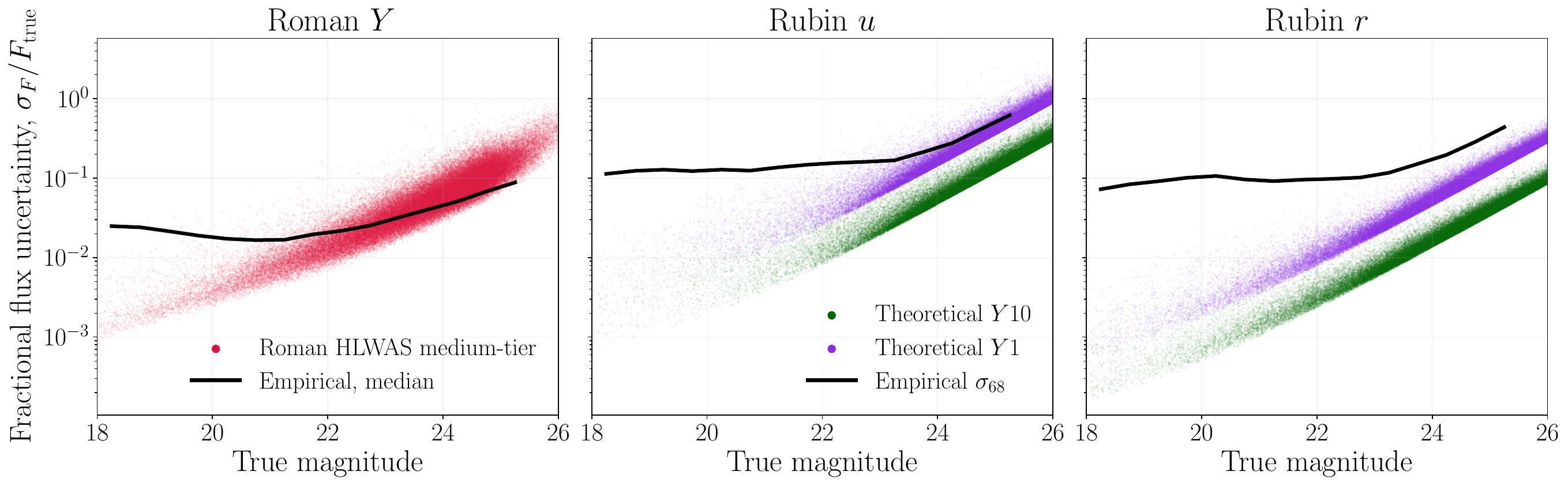}
    \caption{The theoretical fractional flux errors for galaxies as calculated in Sec.~\ref{subsec:error_theoretical}, as a function of true magnitude, for the Roman Y and Rubin $u,r$ bands (left: Y, middle: $u$, right: $r$). The black curves represent the median flux uncertainty from the empirical error modeling (Sec.~\ref{subsec:error_training}) with LSST-Y1 depth. The other Roman bands exhibit a similar trend as the Y band plotted here, and for the Rubin bands, we select bands with the two extreme results in terms of the agreement between the theoretical error and the empirical error (violet: theoretical LSST-Y1, green: theoretical LSST-Y10). Note that for the Y and \textit{u} band, the empirical errors (black lines) asymptote to the theoretical errors (colored dots) on the fainter end, which confirms the validity of our theoretical error calculation (Sec.~\ref{subsec:error_theoretical}), while for \textit{r} band, there remains a systematic error floor.}
    \label{fig:theory_error}
\end{figure*}

We convert this point-source flux noise into that of extended sources (galaxies in this case) by applying a galaxy-size correction to the effective area ($N_{\rm eff}$). We first calculate the half-light radius of the PSFs ($r_{\rm 50,PSF}$). For Roman, the PSFs in the coadded images are Gaussian with the FWHM of 0.24$\arcsec$ (Sec.~\ref{subsec:RomanIMCOM}), for which the half-light radius is theoretically known as ${\rm FWHM}/2$. For Rubin images, we approximate the PSF with a Moffat profile with $\beta=3.5$ \citep[this is a commonly used value for ground-based telescopes;][]{Chang2015,Sheldon2017,Zhang2022}. The corresponding $r_{\rm 50,PSF}$ is also theoretically known as $({\rm FWHM}/2)\sqrt{(2^{1/(\beta-1)} - 1)/(2^{(1/\beta)} - 1)}$. 

From the truth catalogs (Sec.~\ref{subsec:diffsky}), we then numerically calculate the intrinsic half-light radii of the galaxies ($r_{\rm 50,gal}$). As in the OU2024 simulation, we use a Sérsic profile with $n=1$ for the disk and $n=4$ for the bulge. Then using the disk fraction in the catalog, we make the composite disk+bulge profiles for all galaxies, and numerically calculate the corresponding $r_{\rm 50,gal}$ values.

We then apply the galaxy-size correction to $V_{\rm point,PSF}$ to obtain the final noise variance of galaxies $V_{\rm ext,PSF}$:
\begin{equation}
\label{eq:galaxy_size_correction}
    V_{\rm ext,PSF} = V_{\rm point,PSF} \left(\frac{r_{\rm 50,PSF}^2 + r_{\rm 50,gal}^2}{r_{\rm 50,PSF}^2}\right)
\end{equation}
to account for the change in the effective area that galaxies occupy. Note that this equation assumes a Gaussian galaxy light profile, while the realized galaxy images follow a composite Sérsic profile with $n=1$ and $n=4$. While we acknowledge that it can affect the accuracy of this equation, we defer a more detailed study on this to future studies. Including the self-Poisson noise from a galaxy of flux $F_{\rm gal}$, the final theoretical flux error of each galaxy becomes,
\begin{equation}
    V_{\rm F,gal} = V_{\rm ext,PSF} + F_{\rm gal}.
\end{equation}
Note that we use two LSST survey years for this analysis for comparison: Y1 and Y10. When assigning actual flux errors to each galaxy, we randomly draw an error value from a Gaussian distribution with the mean of zero and the variance of $V_{\rm F,gal}$.

We then empirically define the 10$\sigma$ limiting magnitude ($m_{10}$) by selecting galaxies with SNR between 9.995 and 10.005 and calculating the median true magnitude for each band. We apply the corresponding $m_{\rm gal,H} < m_{\rm 10,H}$ selection to the new \texttt{Diffsky} mock galaxy catalog (Sec.~\ref{subsec:diffsky}) to construct our final sample for the subsequent photo-$z$ analysis. Note that we apply this selection consistently for the galaxy sample with the theoretical error calculation in this section and that with the empirical error modeling in Sec.~\ref{subsec:error_training} for fair comparison.

In Fig.~\ref{fig:theory_error}, we plot the fractional flux uncertainties from the theoretical error calculation and the empirical error modeling. We plot the Roman Y band, and Rubin \textit{u, r} bands as representative samples. On the left panel, we plot the calculated flux uncertainty for the Roman Y band as a function of true magnitude. The red dots show the theoretical errors in this section and the black solid line represents the median of the empirical errors modeled in Sec.~\ref{subsec:error_training} for comparison. On the brighter end, as expected from Sec.~\ref{subsec:photometry_result_roman}, the empirical errors show a significantly higher level of flux uncertainty than the theoretical (statistical) errors, implying a systematic uncertainty floor (a flat curve below the magnitude of $\sim$22). On the other hand, towards the fainter end, we confirm that the two calculations asymptote to each other, confirming the validity of our theoretical error calculation, as statistical error becomes dominant in the total error budget for faint galaxies. However, note that for this Roman Y-band, the faint end slope does not agree very well between the empirical and theoretical cases. This could indicate additional systematic effects in the error properties, such as the significant correlated noise described in Sec.~\ref{subsec:correlated_noise}, or due to the effect of the greater depth of Roman than Rubin.

The result for Rubin \textit{u} band (middle panel) closely follows that of the Roman Y band, where the theoretical errors with the LSST-Y1 depth (violet dots) and the empirical errors (also with the LSST-Y1 configuration, black) asymptote to each other on the fainter end, while the empirical curve becomes flat below magnitude of $\sim$22.5 (a systematics floor). Also note that the calculated theoretical errors with the LSST-Y10 depth (green dots) are on average $\sim$3 times smaller than the Y1 errors, as expected. 

Rubin $r$ band, chosen as an extreme case example among the Rubin bands, shows a somewhat different behavior. While showing a similar trend to that of \textit{u} band, the empirical errors (black) show a systematically higher level of uncertainty than the theoretical errors. Note that our theoretical error calculation implicitly assumes Gaussian profiles for PSFs and galaxy shapes (quadrature summation), from which the effective area is calculated to account for the PSF and galaxy size (Eq.~\ref{eq:galaxy_size_correction}). However, our empirical errors are based on the GAaP photometry on simulated images where the measured PSF size and the effective area for a GAaP aperture can be different from those of the theoretical calculation. Also, while the $r$ band is theoretically expected to have the smallest error level among the Rubin bands, the empirically measured flux residuals for the $r$ band show even slightly larger scatter than the $g$ band. This combination makes the $r$-band discrepancy especially pronounced. Given that these effects could generally make a discrepancy in scatter between empirical and theoretical flux residuals, the faint-end discrepancy between them for some Rubin bands is not completely unexpected.

\section{Impact on photometric redshift estimation}
\label{sec:photoz}

With the new \texttt{Diffsky} mock galaxy catalog assigned the theoretical and empirical flux uncertainties from the previous section, we perform a set of photo-$z$ analyses in this section. We aim to investigate the effect of 1) Rubin depths, 2) systematic uncertainties in photometry, and 3) different survey combinations, on the Roman--Rubin combined photo-$z$ estimation performance. Note that, in this study, we use two Roman--Rubin combinations: Roman-LSST-Y10, and Roman-LSST-Y1. While Roman-LSST-Y10 is not a contemporaneous combination, it represents an idealized upper limit for the synergy between the two surveys.

\subsection{Photometric redshift estimation algorithms}
\label{subsec:photoz_algorithm}

We use the Redshift Assessment Infrastructure Layers \citep[RAIL\footnote{https://github.com/LSSTDESC/RAIL},][]{RAIL}  package to estimate the photo-$z$s of our simulated galaxies (with $m_{\rm gal,H} < m_{\rm 10,H}$ and $0<z<3$ selection). RAIL, originated by the LSST Dark Energy Science Collaboration (LSST-DESC), is a software library that provides efficient tools for producing photo-$z$ data products from large data sets, including uncertainties and summary statistics, and functionality to stress-test the results under realistically complex systematics. We refer readers to \citet{RAIL} for more details.

We use two of the photo-$z$ estimation algorithms implemented in RAIL: \texttt{FlexZBoost} \citep{FZBoost, FZBoost2, RAIL} and \texttt{LePHARE} \citep{Arnouts1999,Ilbert2006}. The former uses a series of machine-learning algorithms to optimize the photo-$z$ estimation process, and the latter uses an SED template-fitting method with an extensive set of SED templates. We choose \texttt{FlexZBoost} and \texttt{LePHARE} as representative algorithms for the machine-learning-based and template-fitting methods, respectively, which are the two main branches of modern photo-$z$ estimation methods. Also, these two algorithms exhibited the best results for each branch within RAIL when applied to Rubin Data Preview 1 \citep{dp1_photoz}. While we refer readers to these references, we briefly describe their algorithms here. 

\texttt{FlexZBoost} is a supervised photo-$z$ estimation algorithm that predicts the conditional redshift probability density, $p(z|\boldsymbol{x})$, where $\boldsymbol{x}$ contains magnitudes (colors) and photometric uncertainties. Based on FlexCode \citep{FZBoost, FZBoost2}, it uses gradient-boosted decision trees to learn how galaxies with similar photometric properties are distributed in the redshift space. The conditional density, $p(z|\boldsymbol{x})$, is represented by a linear combination of orthonormal bases (cosine or Fourier):
\begin{equation}
    p(z|\boldsymbol{x}) = \sum_k \beta_k(\boldsymbol{x}) \phi_k (z),
\end{equation}
where $\phi_k (z)$ are the basis functions, and $\beta_k(\boldsymbol{x})$ are the corresponding coefficients. The algorithm uses the \texttt{XGBoost}\footnote{https://xgboost.readthedocs.io/en/stable/} implementation of GBDT to predict the coefficients, $\beta_k (\boldsymbol{x})$, given the photometry of the training sample. The trained coefficients are then combined to construct the final estimation of $p(z|\boldsymbol{x})$. We use the cosine basis functions with the maximum number of bases of 35. We split the sample into a 75\% training and a 25\% validation sample. The bump threshold parameter\footnote{It removes low-amplitude likely spurious peaks in $p(z)$.} is selected from 20 values from 0.02 to 0.35, and sharpening parameter\footnote{It adjusts the overall concentration of $p(z)$.} among 15 values from 0.7 to 2.1. For easier comparison between different survey configurations, we use a training sample representative of the entire sample.

\texttt{LePHARE} is a template-fitting-based photo-$z$ estimation algorithm using a set of SED templates. Since we are only interested in galaxy redshift, we only use the galaxy SED templates based on the COSMOS library \citep{Ilbert2009}. For each galaxy at each redshift, it optimizes the template normalization and evaluates the $\chi^2$ of the photometry between the data and the model, while allowing dust attenuation and empirical emission line contributions. The resultant likelihood is then combined over all templates to construct the final prediction of $p(z)$. 

In the following analyses, we apply the above photo-$z$ estimation algorithms to the new \texttt{Diffsky} mock catalog (Sec.~\ref{subsec:diffsky}) to which photometric noise has been added according to the theoretical errors (Sec.~\ref{subsec:error_theoretical}) and the empirical errors using the emulator from Sec.~\ref{subsec:error_training}. We use these analyses to assess the effects of photometry biases and uncertainties, as well as different survey combinations, on photo-$z$ performance.

\subsection{Quality of photometric redshift estimation as a function of Rubin depth}
\label{subsec:photoz_theoretical_error}

In this subsection, we focus on the impact of Rubin depths on the photo-$z$ performance with the combined Roman--Rubin configuration using the theoretical error model. For this analysis, we also construct a ``perfect'' Rubin-Roman photometry sample, in which the true magnitudes are assigned to each galaxy while setting the magnitude error at $10^{-8}$. Since \texttt{LePHARE} uses $\chi^2$ statistics and these extremely small magnitude errors result in failures in fitting, we only run this perfect case with \texttt{FlexZBoost}. In this section, we compare three survey configurations: perfect Roman-Rubin photometry, theoretical Roman-LSST-Y10 errors, and theoretical Roman-LSST-Y1 errors to compare how different survey depths affect the photo-$z$ estimation. 

In this analysis, we report three summary statistics for the photo-$z$ performance, based on the values of $\Delta z \equiv (z_{\rm true} - z_{\rm phot})/(1 + z_{\rm true})$, where $z_{\rm true}$ is the true redshift, and $z_{\rm phot}$ is the estimated photo-$z$: 
\begin{itemize}
    \item The robust NMAD (normalized median absolute deviation; $\sigma_{\rm NMAD}$): we iteratively calculate the NMAD of a photo-$z$ sample by excluding the catastrophic outliers with $|\Delta z - {\rm median}(\Delta z)| > 5 \sigma_{\rm NMAD} (\Delta z)$, and in turn calculating the NMAD again. We repeat this process until no additional catastrophic outlier is detected. 
    \item The robust bias ($b_{\rm z}$): using the converged sample from the above iterative process, we calculate the median of $\Delta z$,
    \item The outlier fraction ($f_{\rm z,out}$): we identify photo-$z$ outliers by tagging galaxies with $|\Delta z| > 0.15$, i.e., the photo-$z$ error exceeding 15\%, as outliers \citep{dp1_photoz}. Note that we use an absolute threshold of 0.15 for the outlier fraction instead of the catastrophic outlier definition as part of the above iterative NMAD calculation, to prevent the outlier fraction from being strongly dependent on the value of NMAD itself. 
\end{itemize}
From now on, in this paper, ``bias'' represents the robust bias, and ``NMAD'' represents the robust NMAD. Note that while this study focuses on a comparative (rather than absolute) analysis regarding the impact of photometry performance on photo-$z$ estimation, we remind readers that the Roman Science Requirements Document \citep{RomanSRD} requires 1$\sigma$ (68\%) ranges of $p(z)$'s of 50\%, 30\%, and 20\% of galaxies to be within 0.04, 0.08, 0.15 in $\Delta z$, respectively, while \citetalias{Rotac} predicts that the Roman--LSST-Y4 combination would exhibit the NMAD of 0.032 with a note that this number is likely to be overly optimistic.

\begin{figure*}
    \includegraphics[width=\textwidth]{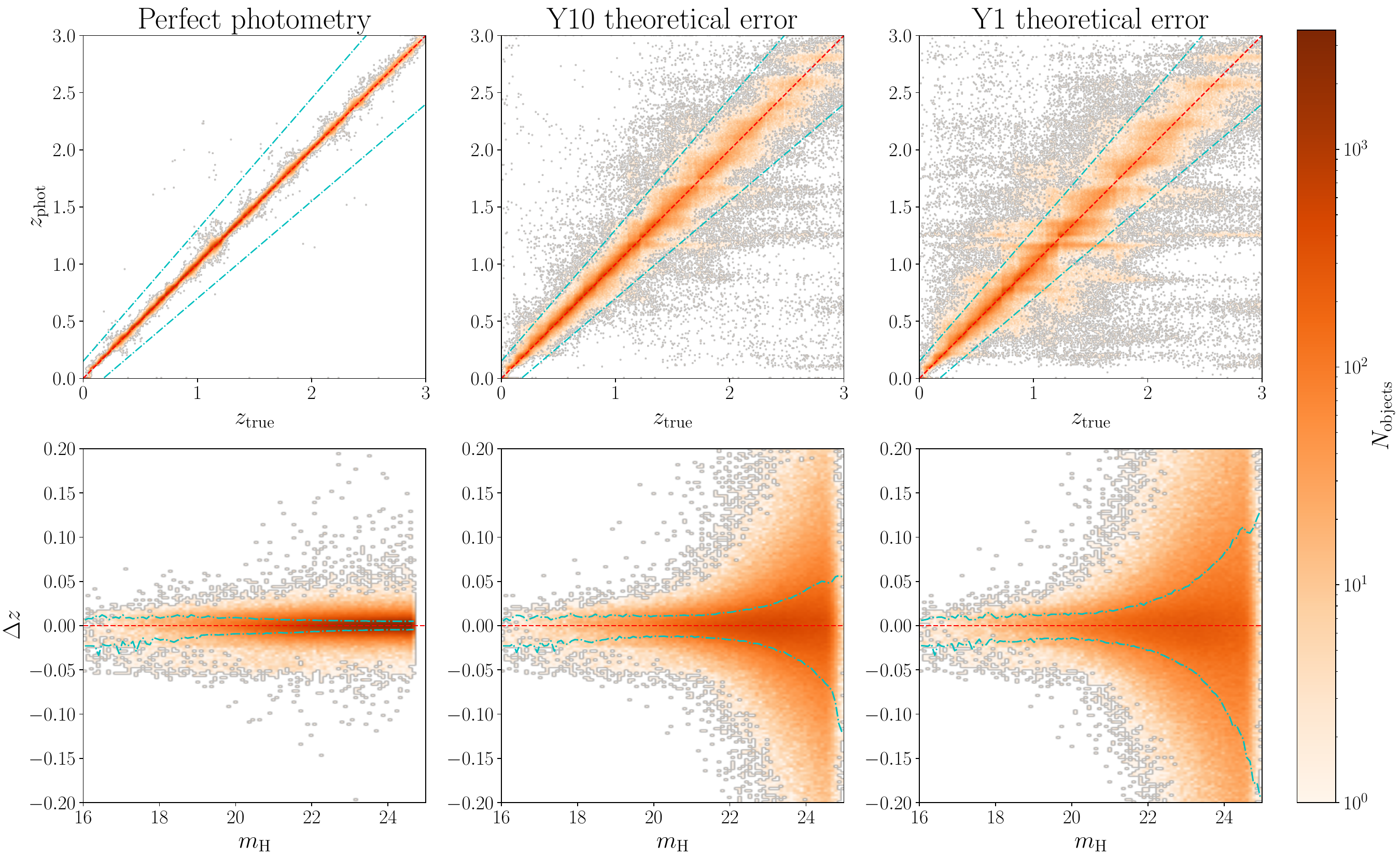}
    \caption{\textit{top}: true redshift vs.\ photometric redshift from the perfect Rubin + Roman photometry (using true magnitudes with magnitude errors of $10^{-8}$ for all galaxies and bands; left), LSST-Y10 + Roman HLWAS medium tier theoretical errors (middle), and LSST-Y1 + Roman HLWAS medium tier theoretical errors (right). The cyan lines correspond to our outlier threshold ($|\Delta z| >0.15$). \textit{bottom}: photo-$z$ error as a function of Roman H-band magnitude. The cyan lines denote the 68\% confidence interval.}
    \label{fig:ztrue_zphot}
\end{figure*}

The left panels of Fig.~\ref{fig:ztrue_zphot} show the true redshifts compared to the photo-$z$s with the perfect Roman-Rubin configuration. The top panel shows the relation between $z_{\rm true}$ and $z_{\rm phot}$, and the bottom panel shows the photo-$z$ bias as a function of true H magnitude. The photo-$z$ and the true redshift closely follow the one-to-one relation as expected and there is not any apparent magnitude trend in the photo-$z$ performance. The outlier fraction is 0.01\%, the bias is 3.4$\times$$10^{-5}$, and the NMAD is 0.005, indeed showing almost perfect photo-$z$ performance and therefore confirming the validity of our photo-$z$ pipeline. 

In the middle (right) panels in Fig.~\ref{fig:ztrue_zphot}, we plot the FlexZBoost result using the combination of Roman HLWAS medium tier, which we refer to as ``Roman'' hereafter, and the LSST-Y10 (Y1) with theoretical errors. In the upper panels, cyan lines show our outlier threshold of $|\Delta z| > 0.15$, and the cyan lines in the bottom panels show the 68\% confidence intervals. For the Roman-LSST-Y10 case, the outlier fraction increases to 3.6\%, while the NMAD and the bias also increase to 0.022 and -5.8$\times$$10^{-4}$, respectively, compared to the perfect case. However, the photo-$z$s still exhibit a good agreement with the true redshift at all redshifts (upper panel). 

In the perfect photometry case, $\Delta z$ does not show any apparent magnitude trend, because the magnitude errors are independent of the magnitudes. However, since magnitude errors increase with magnitude in realistic photometry (Fig.~\ref{fig:theory_error}), we observe increasing photo-$z$ scatter towards the fainter end (bottom middle panel) as one would expect for this theoretical Roman-LSST-Y10 case.\footnote{Note that with our representative training sample, the photo-$z$ performance in this study will be overly optimistic compared to real data where one usually does not have a representative spectroscopic training sample. However, since the purpose of this study is to perform a comparative study under an ideal condition, this is not a limiting factor for our analysis.}

The Roman-LSST-Y1 case also shows a similar trend as the Roman-LSST-Y10 case, but with increased scatter compared to the LSST-Y10 case. The measured outlier fraction is 11.5\%, and the NMAD is 0.036: a 220\% and 66\% increase, respectively. The bias of -1.4$\times$$10^{-3}$ does not show any meaningful difference compared to the LSST-Y10 case. Note that the redshift degeneracy (the $z_{\rm true}$ distribution given $z_{\rm phot}$) is more clearly seen in this configuration (horizontal strips in the top right panel). 

\begin{figure*}
    \includegraphics[width=\textwidth]{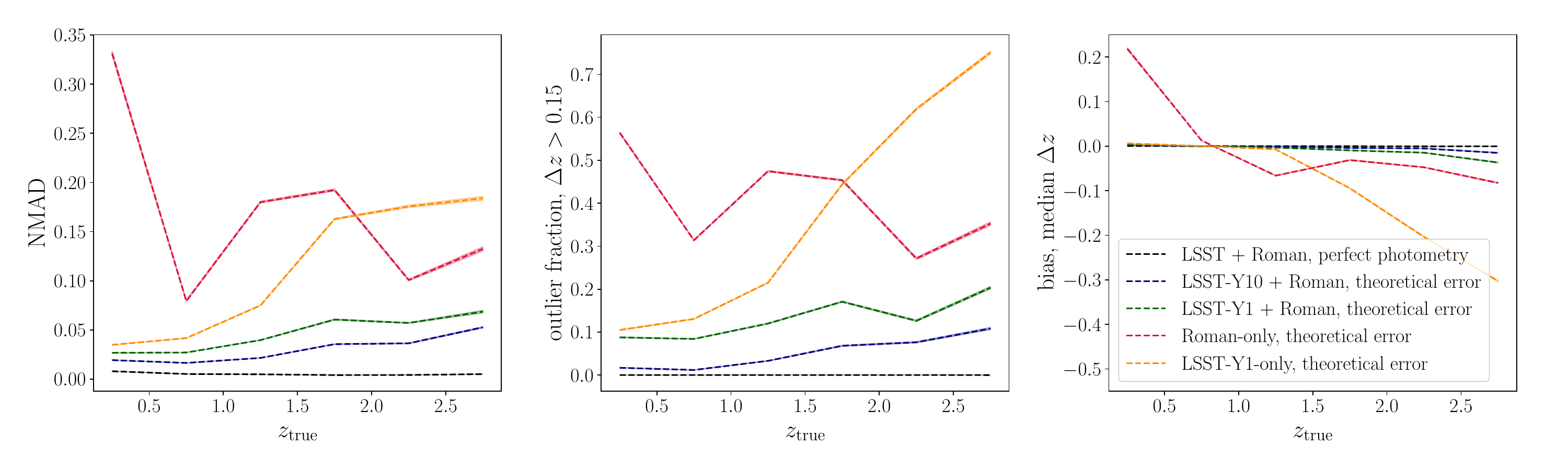}
    \caption{The normalized median absolute deviation (NMAD, left), the outlier fraction ($>15$\% deviation, middle), and the median bias (right) of the photometric redshift estimates with FlexZBoost algorithm on the perfect LSST + Roman photometry (black), LSST-Y10 + Roman HLWAS medium tier theoretical photometry errors (blue), LSST-Y1 + Roman HLWAS medium tier theoretical photometry errors (green), Roman HLWAS medium tier only theoretical photometry errors (red), and LSST-Y1 only theoretical photometry errors (yellow). See Sec.~\ref{subsec:error_theoretical} for more information.}
    \label{fig:photo-$z$_theoretical}
\end{figure*}

In Fig.~\ref{fig:photo-$z$_theoretical}, we plot the three summary statistics (NMAD -- left, outlier fraction -- middle, bias -- right) as a function of redshift (6 bins between 0 and 3), for the three configurations we have discussed in this section (perfect case -- black, Roman-LSST-Y10 -- blue, Roman-LSST-Y1 -- green). First, we observe that, as photometry errors increase with redshift (fainter galaxies), the NMAD scatter and the outlier fraction also increase with redshift for all three configurations. For the NMAD scatter and the outlier fraction, we confirm the expected trend that the Roman-LSST-Y10 combination (blue) performs better than the Roman-LSST-Y1 combination (green), due to its higher depth. However, none of the three configurations shows a meaningful bias across the entire redshift range, while there is a hint of deteriorating bias performance for the Roman-LSST-Y1 at the highest redshift bin.

\begin{figure*}
    \includegraphics[width=\textwidth]{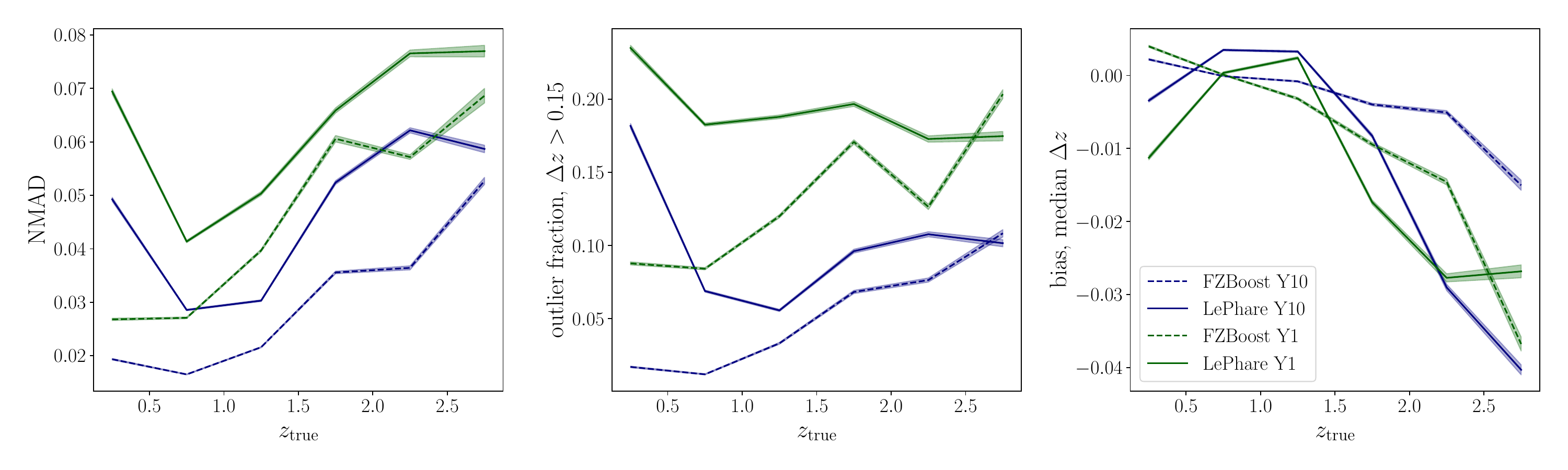}
    \caption{Similar to Fig.~\ref{fig:photo-$z$_theoretical}, but comparing \texttt{FlexZBoost} (dashed) with \texttt{LePHARE} (solid) with LSST-Y10 (Y1) + Roman HLWAS medium tier theoretical photometry errors in blue (green).}
    \label{fig:photo-$z$_compare}
\end{figure*}

Thus far in this section, we have quantified how different the photo-$z$ performance would be when combining Roman with LSST-Y10 and LSST-Y1 using \texttt{FlexZBoost} algorithm. As \texttt{FlexZBoost} is a machine-learning based algorithm, it would be informative to compare its performance to a template-fitting photo-$z$ algorithm. Therefore, we qualitatively compare the result using \texttt{FlexZBoost} with that using \texttt{LePHARE}, for these two combinations: the theoretical Roman-LSST-Y10 and Roman-LSST-Y1 case. Note that we do not quantify the difference between the results from these two algorithms, as different software configurations could significantly alter the results.

In Fig.~\ref{fig:photo-$z$_compare}, we plot the three summary statistics for \texttt{FlexZBoost} and \texttt{LePHARE}. The solid lines represent the results with \texttt{LePHARE} and the dashed lines those with \texttt{FlexZBoost}. The blue color corresponds to the Roman-LSST-Y10 combination, and the green corresponds to the Roman-LSST-Y1 combination. For the NMAD scatter and bias, both algorithms show similar trends: increasing scatter and bias towards higher redshifts. However, while \texttt{FlexZBoost} exhibits an increasing fraction of photo-$z$ outlier towards higher redshifts, \texttt{LePHARE} shows a fairly steady level of outlier fraction across all redshifts. Also, the trend between the LSST-Y1 and LSST-Y10 is qualitatively preserved across the two algorithms. 

We note that \texttt{LePHARE} could suffer from the incomplete template library compared with the input SEDs of our simulated galaxies, while the detailed configuration for running the algorithm might strongly affect the photo-$z$ performance. Similarly, \texttt{FlexZBoost} can also suffer from a non-representative training sample. However, given the current level of agreement in the qualitative photo-$z$ trend and the convenience of \texttt{FlexZBoost} to be able to deal with very small flux errors, we decide to focus on the \texttt{FlexZBoost} algorithm hereafter.

\subsection{Impact of systematic uncertainties in photometry on photo-$z$ estimation}
\label{subsec:photoz_empirical_error}

\begin{figure*}
    \includegraphics[width=\textwidth]{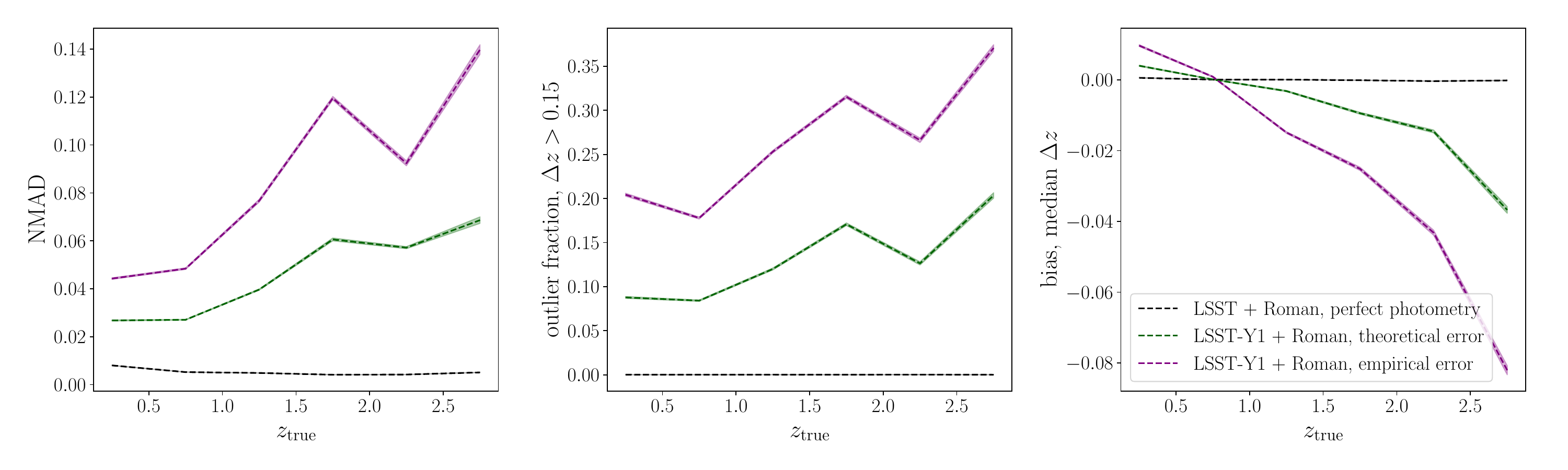}
    \caption{Similar to Fig.~\ref{fig:photo-$z$_theoretical}, but on the perfect LSST + Roman photometry (black), LSST-Y1 + Roman HLWAS medium tier theoretical photometry errors (green), and LSST-Y1 + Roman HLWAS medium tier empirical errors from the LSST Science Pipelines (violet). See Sec.~\ref{sec:error_modeling} for more information.}
    \label{fig:photo-$z$_theoretical_vs_empirical}
\end{figure*}

Next, we compare the photo-$z$ estimation performance with the empirical error modeling (Sec.~\ref{subsec:error_training}) to that with the theoretical errors (Sec.~\ref{subsec:error_theoretical}) and that with the perfect photometry, using the Roman-LSST-Y1 configuration to test how additional systematic uncertainties in realistic photometry affect the photo-$z$ estimation.  

In Fig.~\ref{fig:photo-$z$_theoretical_vs_empirical}, we plot the NMAD (left), the outlier fraction (middle), and the bias (right) as a function of true redshift for these three configurations. The black (perfect photometry) and green (theoretical Roman-LSST-Y1 errors) curves are identical to those in Fig.~\ref{fig:photo-$z$_theoretical} with the same colors. When compared to the case with the theoretical, statistical-only errors (green), the empirical error model, which includes various systematic uncertainties (e.g., centroid offset, model mismatch, blending, etc.), exhibits significantly larger scatter and outlier fraction. The ensemble NMAD scatter for the case of empirical errors is 0.065 compared to 0.036 for the theoretical error case, and the corresponding outlier fraction is 24\% compared to 11.5\% for the theoretical case. While we do not know whether the degradation in photo-$z$ performance is additive or multiplicative, both statistics show approximately 2 times worse results for the empirical case compared to the theoretical case. In addition, the corresponding bias is -0.006 while that for the theoretical error case is -0.0014, also showing a significantly less optimal result for the empirical case.

The overall redshift trend remains qualitatively the same as the previous section: a downtrend in the photo-$z$ performance with increasing redshift. For the NMAD (outlier fraction) the empirical case shows $\sim$0.027 scatter ($\sim$20\% outliers) at $0<z<0.5$, while at $2.5<z<3$ it exhibits $\sim$0.069 scatter ($\sim$37\% outliers). Also, at the highest redshift bin, the bias for the empirical case reaches -0.08, which is not negligible. On the other hand, the relative degradation from the theoretical case (green) to the empirical case (violet) remain stable across all redshift values, at approximately 2 times for the NMAD scatter and the outlier fraction. Thus, although the absolute photo-$z$ performance worsens toward higher redshifts, the additional photometry effects captured by the empirical model do not appear to become disproportionately more severe with redshift. Reducing these systematic effects remains important across the full redshift range, rather than only for faint, high-redshift galaxies.

\subsection{Photometric redshift with different survey combinations}
\label{subsec:photoz_survey_combination}

\begin{figure*}
    \includegraphics[width=\textwidth]{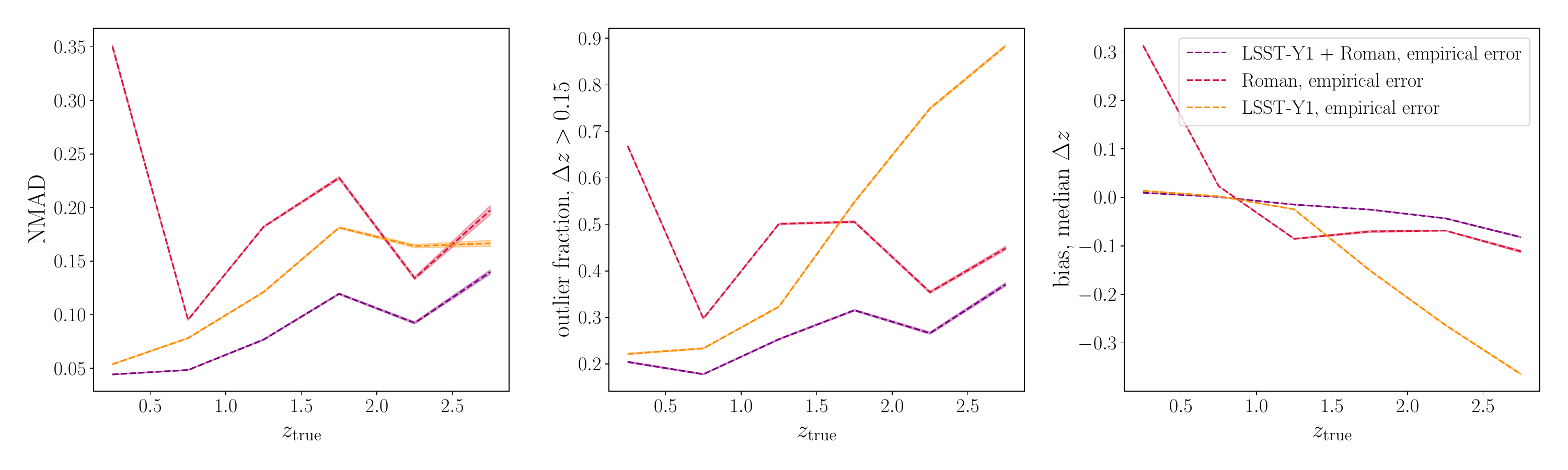}
    \caption{Similar to Fig.~\ref{fig:photo-$z$_theoretical}, but on the LSST-Y1 + Roman HLWAS medium tier (violet), Roman medium tier only (red), and LSST-Y1 only (orange) with the empirical errors from the LSST Science Pipelines. See Sec.~\ref{sec:error_modeling} for more information.}
    \label{fig:photo-$z$_roman_rubin_romanrubin}
\end{figure*}

We also evaluate the effect of different survey combinations on the photo-$z$ performance. We compare three cases, Roman-LSST-Y1, Roman-only, and LSST-Y1-only, with both the theoretical errors and the empirical error modeling. 

In Fig.~\ref{fig:photo-$z$_theoretical}, we show the summary statistics for these three cases with the theoretical errors. The green lines represent the Roman-LSST-Y1 case, the red lines the Roman-only case, and the yellow lines the LSST-Y1-only case. Note that for the LSST-Y1, the selection is still made in the Roman H band, therefore the result could be different from results for a Rubin-selected sample. 

For the Roman-only case, the scatter and the outlier fraction are very high at low to intermediate redshift, and tend to decrease (improve) with redshift (also note the high bias value of $\sim$0.2 at the lowest redshift bin). This is due to the fact that, as the 4000 Å break enters the infrared band around $z$$\sim$1.5, the Roman bands ($\sim$1-2 $\mu$m) carry more information towards higher redshift above it. This effect becomes clearer when we compare the LSST-Y1-only curves (yellow) to the Roman-LSST-Y1 curves (green). The LSST-Y1-only summary statistics (NMAD, outlier fraction, bias) abruptly deteriorate above $z$$\sim$1.5: the NMAD (outlier fraction) is $\sim$0.03 (10\%) for the lowest redshift bin, increasing to $\sim$0.18 (75\%) for the highest redshift bin. The photo-$z$ bias for the LSST-Y1-only case also significantly deviates from 0 at $z$$>$1.5, down to -0.3 for the highest redshift bin. However, when we combine the LSST-Y1 with Roman, the summary statistics become stable across all redshifts, showing much improved photo-$z$ performance (below 0.07 for NMAD, below 20\% for outlier fraction, below 0.04 for the bias, across all redshifts). Also note that above $z$$\sim$1.5, the Roman-only case outperforms the LSST-Y1-only case. 

While combining the Roman and Rubin bands improves the Rubin photo-$z$ mainly at high redshifts ($z$$>$1.5, where the improvement is marginal below it), adding the Rubin bands to the Roman bands has a significant impact over the entire redshift range, improving the NMAD (outlier fraction) by $\sim$45-90 ($\sim$45-85)\% across the redshift range. Together, our findings are that 1) the Roman photometry carries significant information especially above the $z\sim$1.5, which can complement the Rubin photo-$z$, while 2) including the Rubin bands significantly improves the Roman photo-$z$ at all redshifts.

In Fig.~\ref{fig:photo-$z$_roman_rubin_romanrubin}, the photo-$z$ summary statistics with the empirical error modeling (Sec.~\ref{subsec:error_training}) for these three survey combinations are shown. While they exhibit similar trends to those in the above theoretical error case, due to the increased systematic uncertainty budget, they show deteriorated photo-$z$ performance compared with the theoretical case. For the Roman-LSST-Y1 case (violet), the NMAD (outlier fraction) increases by 60-100\% (20-110\%) at all redshifts. The bias values also increase by 2 to 9 times. For the Roman-only case (red)), the NMAD scatter, the outlier fraction, and the bias increase up to 50\%, 30\% and 80\%, respectively. However, the LSST-Y1-only case shows an interesting trend; while the outlier fraction and the bias exhibit an expected 20-110\% and 20-480\% increase at all redshifts, respectively, the NMAD scatter shows similar values to those of the theoretical case above $z=1.5$ within $\pm$10\% (50-90\% increase below $z=1.5$). This implies that above $z=1.5$, the photo-$z$ distribution is systematically biased low with the scatter changing minimally, which is also consistent with the fact that the Rubin-only configuration provides progressively less redshift information above $z=1.5$. Relatedly, note that for this empirical case, even the Roman-LSST-Y1 combined photo-$z$ result exhibits a significant bias level above $z$$\sim$1.5, up to $\sim$8 percent at the highest redshift bin. Together, this implies that the rich information that the Roman photometry carries above $z$$\sim$1.5 gets significantly diluted due to the systematic errors of LSP photometry. Together, these findings underscore the need for further optimization of the LSP photometry pipelines for the future Roman-Rubin combined analysis. 

Note that, given the similar performance of color retrieval between \texttt{Metadetect pgauss} and LSP GAaP/CModel (Fig.~\ref{fig:color_residual_compare}), we expect the corresponding photo-$z$ performance to be similar as well to each other. Since running Rubin photometry using \texttt{Metadetection} pipeline is beyond the scope of this study, we defer a quantitative analysis to a future study.

\section{Summary and discussion}
\label{sec:conclusion}

In this study, we have conducted a photometry analysis using the LSST Science Pipelines (LSP) on simulated Rubin and Roman coadded images using the Roman H band as the reference band and performing forced photometry on all the other bands. Then we assessed the photometry performance by comparing the results against the truth. We have also compared it with the \texttt{pgauss} photometry algorithm as part of the \texttt{Metadetection} shear measurement pipeline. Since the Roman \texttt{IMCOM} coadded images retain highly correlated image noise across several pixels, we have also calculated the flux variance of each measured galaxy accounting for the correlated noise (for the Roman bands with LSP) and compared it with the uncertainty obtained from the diagonal (uncorrelated) noise only. Our key findings regarding photometry are as follows:
\begin{itemize}
    \item Roman CModel photometry is on average unbiased over most of the flux range. However, the fractional flux scatter reaches a systematic floor of $\sim$7-8\% for bright galaxies.
    \item LSP CModel and \texttt{Metadetection pgauss} differ strongly in absolute flux recovery (\texttt{pgauss} significantly underestimates total flux especially for bright galaxies).  However, after median subtraction, both reach a systematic flux uncertainty floor of 7-8\%, implying that neither is purely dominated by statistical uncertainty, hinting at a similar systematic cause. Tests with isolated galaxies indicate that blending is not the dominant contributor to this systematics floor.
    \item Despite the significant difference in flux bias and scatter between LSP CModel and \texttt{pgauss}, they recover very similar color distributions. Interestingly, the color scatter becomes negligible at the bright-end, unlike for the flux recovery, implying that the bright-end flux systematics are strongly correlated across bands.
    \item Due to their algorithmic similarity, LSP GAaP shows a similar behavior as \texttt{Metadetection pgauss} in both flux bias and color estimation, with GAaP showing somewhat smaller scatter. This supports our choice to use GAaP for estimating galaxy colors.
    \item Accounting for correlated pixel noise is essential for realistic flux errors. Ignoring correlated noise results in the flux error being underestimated by a factor of $\sim$4, which affects SNR-based selection.   
    \item Rubin forced photometry is approximately unbiased in color overall, with \textit{$u-g$} as the main exception showing a pronounced blue tail and larger scatter due to the high fraction of negative \textit{u} band fluxes and weaker \textit{u--g} error correlation. 
\end{itemize}

Using the measured photometry, we then built and validated an empirical machine-learning model that maps the true galaxy properties (true magnitudes, sizes of the disk/bulge, and disk fraction) to the measured fractional flux error, incorporating realistic biases and scatter in the photometry including correlations across bands, without parametric assumptions about the form of the scatter. 
We applied this model to the new \texttt{Diffsky} mock galaxy catalog with improved realism in the galaxy SEDs, thereby increasing the realism of both the underlying galaxy model and of the observational noise model used for a subsequent photo-$z$ analysis -- without having to run expensive image simulations and photometry analysis. We also compared the empirical model with the expected statistical (theoretical) flux errors, using a recent Rubin operational simulations database and the reported values in \citetalias{Rotac}, accounting for the size of the point spread function and the size of galaxies. The theoretical error calculation agrees with the empirical machine-learning error model towards the fainter end for the Roman bands as well as for some Rubin bands, supporting the validity of the theoretical error prescription where statistical noise dominates. For the other Rubin bands, we observe discrepancies even towards the faint end, mainly due to the difference in 1) aperture size and 2) relative error properties across bands, between the empirical and theoretical model.

Finally, using these empirical and theoretical error models applied to the new \texttt{Diffsky} mock galaxy catalog, we performed photo-$z$ analyses using different survey and depth configurations with two photo-$z$ estimation algorithms (\texttt{FlexZBoost} and \texttt{LePHARE}). We find:
\begin{itemize}
    \item Rubin survey depth significantly impacts the photo-$z$ precision. With theoretical errors, Roman-LSST-Y10 gives an outlier fraction of 3.6\% and NMAD of 0.022, compared with 11.5\% and 0.036, respectively, for Roman-LSST-Y1. The bias remains comparatively small. The degradation becomes stronger towards faint galaxies and high redshifts.
    \item Applying the empirical error model strongly degrades the photo-$z$ performance compared with the theoretical errors. For Roman-LSST-Y1, the empirical error model increases the ensemble NMAD from 0.036 to 0.065 and the outlier fraction from 11.5\% to 24\%, while the bias changes from -0.0014 to -0.006. The degradation remains stable around a factor of 2 (for the NMAD scatter and outlier fraction) across all redshifts.
    \item Roman and Rubin provide highly complementary wavelength information for photo-$z$ analyses: Roman photometry is particularly important above $z$$\sim$1.5, where Rubin-only photo-$z$ performance steeply deteriorates. Conversely, adding Rubin bands to Roman substantially improves photo-$z$ performance across all redshifts.
    \item Systematic uncertainties in photometry could significantly dilute the Roman-Rubin synergy. With the empirical error model, the combined Roman-LSST-Y1 case exhibits a degraded NMAD, outlier fraction, and bias across the full redshift range compared with the theoretical error case, while reaching a non-negligible photo-$z$ bias of $\sim$8\% in the highest redshift bin. This suggests that reducing systematic photometry errors is important across the full redshift range, particularly for preserving the high-redshift information provided by Roman.
\end{itemize}

There are several remaining questions to be answered in future studies. First, we used somewhat idealized images in this study where we exclude, e.g., background subtraction systematics, bad-pixel masks, foreground object masks, and uncertainties in variance plane reconstruction. These effects could affect the photometry performance and therefore propagate into photo-$z$ estimation in principle. 

We also note several caveats on the Rubin side. In the OpenUniverse2024 simulations, Rubin images are coadded using a traditional coaddition method, which may suffer from edge effects (due to partial coverage from certain single-visit images), the associated inhomogeneity in PSFs \citep{Mandelbaum2023}, and varying depths. The recently released Rubin Data Preview 2 adopted ``cell-based'' coaddition method~\citep{2025rubn.rept...88G}\footnote{also see \url{https://sitcomtn-137.lsst.io/}}, which avoids these issues by only coadding images that cover 100\% of a certain ``cell''~\citep{Armstrong2024}. This cell-based coaddition method became a standard in the Rubin data processing \citep{DP2}. In order to accurately forecast the power of combining Roman and Rubin, Rubin simulations must be processed with the cell-based coaddition method.

In addition, due to limitations of the LSP forced photometry for images with WCS and pixel size that differ from the reference band, only aperture-based photometry methods may be used. To fully optimize the photometry for combining Roman and Rubin data, further software support that enables various photometry choices across different survey configurations may be required.

Also, the current CModel photometry in LSP does not support multi-object fitting. For a deblended parent-children set, it fits each child separately with pixels belonging to other children masked out (single-object fitting). \citet{To2026} showed that for photometry on Rubin coadded images, the single-object fitting method exhibits a non-negligible bias in photometry and photo-$z$. In order to use the LSP to process future Rubin images in combination with those of Roman, further software development may be needed to support various fitting methods.

In this work, we have explicitly tested only one systematic effect in photometry, the effect of blending, while other systematic effects are marginalized in the empirical error modeling (Sec.~\ref{subsec:error_training}) in comparison with the theoretical error (Sec.~\ref{subsec:error_theoretical}). However, there are other important sources of systematics in photometry that require investigation: for example, the effect of centroid offsets, pixel masking, and imperfect background subtraction. In the future, with improved image simulations, we expect to be able to test these systematic effects in detail.

Finally, we have used the Roman H band as a reference band for object detection and selection in this study and performed forced photometry on the other bands. However, there are multiple ways to combine Roman and Rubin photometry, depending on one's specific science goals. For example, one could instead jointly detect and select objects using multiple bands across different surveys, which may further optimize the photometry for combined photometric analyses. Also, applying it to real data, where a reference sample (as truth) is available, will be an important step for further validations of our pipeline. We expect future studies to address this optimization problem for various science cases.

The Nancy Grace Roman Space Telescope will soon collect tens of petabytes of images from space, and the Vera C.\ Rubin Observatory has just started its 10-year LSST program. The combination of Roman’s high-resolution NIR imaging and Rubin’s deep multi-band optical information will make accurate joint photometry and characterization of measurement uncertainties increasingly important for downstream cosmological analyses. Our results provide a basis for understanding how realistic uncertainty in galaxy photometry may affect the expected gains from joint Roman--Rubin photo-$z$ analyses.

\section*{Acknowledgements}

The authors thank Tianqing Zhang, Alex Malz, Brett Andrews, Jeffrey Newman, Philip LaDuca, James Bosch, and Charlotte Ward for helpful discussions.

Part of this work was supported by the NASA ROSES grant 22-ROMAN11-0011, contract number 80NM0024F0012, via a JPL subaward. AK was supported by NASA under award number 80NSSC26K0126.

This work was supported in part by the OpenUniverse effort, which is funded by NASA under JPL Contract Task 70-711320, "Maximizing Science Exploitation of Simulated Cosmological Survey Data Across Surveys". 

This paper makes use of LSST Science Pipelines software developed by the Vera C. Rubin Observatory. We thank the Rubin Observatory for making their code available as free software at https://pipelines.lsst.io. Some of the results in this paper have been derived using the healpy and HEALPix packages.

\section*{Data Availability}

The proprietary data used in this study will be provided upon reasonable request. The OpenUniverse 2024 simulation data including the truth catalog and Rubin images that we used in this analysis is public at \url{https://irsa.ipac.caltech.edu/data/theory/openuniverse2024/overview.html}.


\bibliographystyle{mnras}
\bibliography{references} 



\appendix

\section{The detailed training procedure for the empirical error modeling}
\label{app:diagram}

\begin{figure*}
    \includegraphics[width=\textwidth]{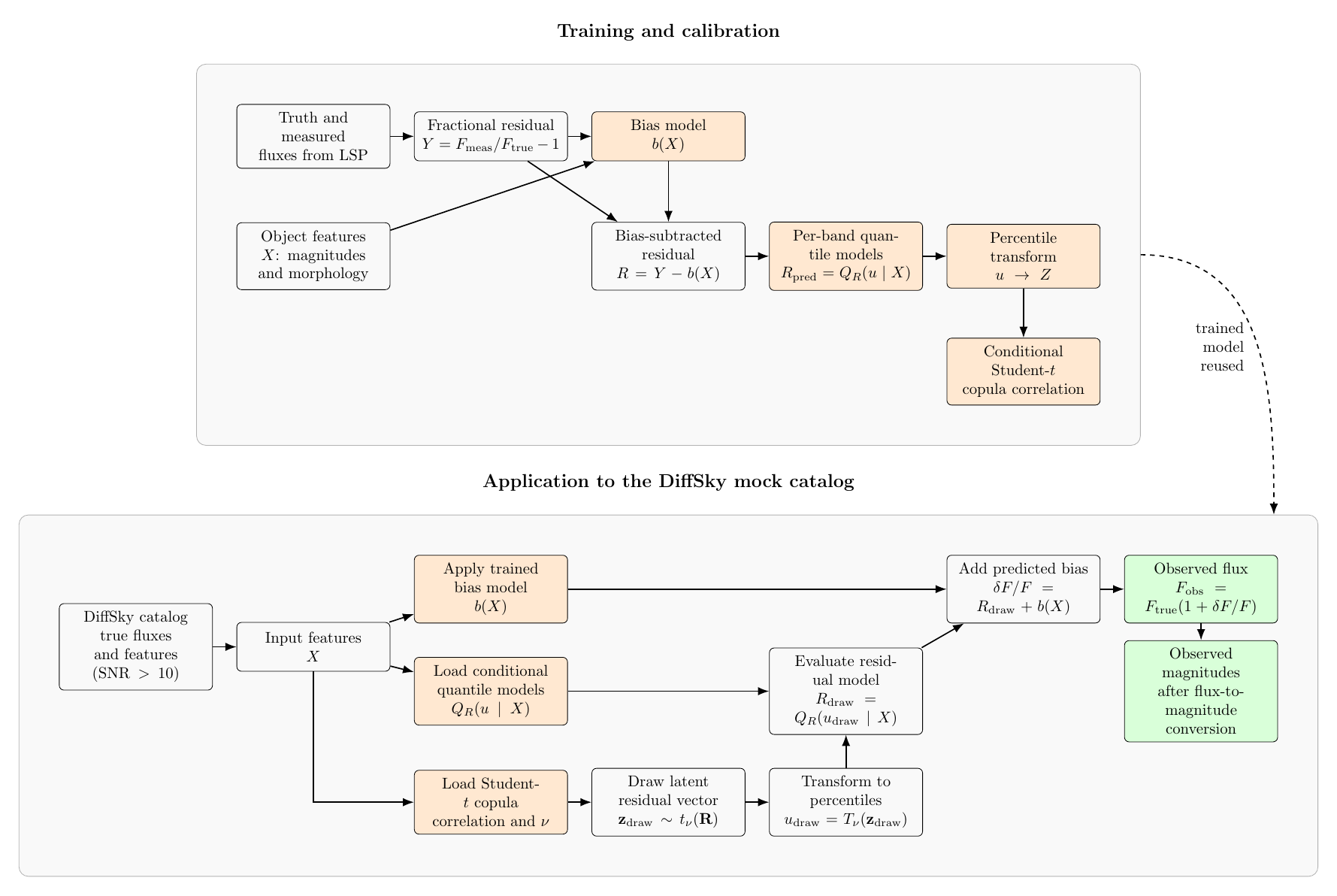}
    \caption{The schematic diagram of our empirical error training algorithm and the corresponding application to a newer version of the Diffsky mock galaxy catalog. The white cells are analysis steps, the yellow cells represent trained models, and the green cells correspond to the final output.}
    \label{fig:diagram_algorithm}
\end{figure*}

The detailed training procedure we outlined in Sec.~\ref{subsec:error_training} is as follows.
\begin{itemize}
    \item We first split the catalog randomly so that 80\% of the galaxies are assigned to the training sample and 20\% to the testing sample.
    \item We train the conditional bias model ($\hat{B}_b(X)$) for every band independently:
    \begin{equation}
        \hat{B}_b(X) = {\rm median}(Y_b | X).
    \end{equation}
    \item We subtract the predicted bias, $\hat{B}_b(X)$, from the fractional flux residual, $Y_b$, to remove the systematic offset:
    \begin{equation}
        R_b = Y_b - \hat{B}_b(X).
    \end{equation}
    \item We train the residual-quantile models. For every band, given a set of 23 discrete percentiles, $\{u\}$, from 0.005 to 0.995\footnote{The used percentiles are 0.005, 0.01, 0.05, 0.1, 0.16, 0.2, 0.25, 0.3, 0.35, 0.4, 0.45, 0.5, 0.55, 0.6, 0.65, 0.7, 0.75, 0.8, 0.84, 0.9, 0.95, 0.99, 0.995.}, we train another set of HGBDTs, \{$\hat{Q}_{R_b}(u|X)$\}: given the input features $X$, it predicts the $u$-th percentile of $R_b$ in each band.
    \item Using these residual-quantile models, we evaluate for each galaxy all 23 values of $R_b$ corresponding to each quantile to obtain the quantile ``grid'': given a galaxy with a property set $X_i$, these are the predicted $u$-th percentile values of $R_b$ for that galaxy.
    \item We take the measured values of $R_b$ in our training sample in each band ($R_{i,b}$) and evaluate their quantiles, $u_{i,b}$, using the interpolated function of the above quantile grid for each respective galaxy:
    \begin{equation}
    \begin{split}
        R_{i,b} &= \hat{Q}_{R_b}(u_{i,b} | X_i) \\
        u_{i,b} &= \hat{Q}_{R_b}^{-1}(R_{i,b}|X_i),
    \end{split}
    \end{equation}
    where the inverse of $\hat{Q}_{R_b}$ is calculated by piecewise-linear interpolation. 
    \item Next, to model the correlation of flux errors between bands, we first transform these estimated percentiles, $u_{i,b}$, into a latent variable, $Z_{i,b}$, which is assumed to follow a Student-\textit{t} distribution\footnote{Note that in statistics, it is called a Student-\textit{t} ``copula''.} with the location parameter 0, the scale parameter 1, and the degrees of freedom $\nu$=4:
    \begin{equation}
        Z_{i,b} = T_\nu^{-1}(u_{i,b}),
    \end{equation}
    where $T_\nu$ is a Student-\textit{t} function with 4 degrees of freedom. This choice gives our model the ability to handle 1) cross-band correlations with the same marginal Student-\textit{t} distribution, 2) simultaneous extreme errors, and 3) heavier joint tails than a Gaussian distribution, with only one additional global parameter, $\nu$ (note that the location parameter and the scale parameter are nuisance parameters that do not change the assumed behavior of $u$).
    \item Then, in the $Z$-space, we calculate the ``local'' correlations between different bands, using the 500 nearest neighbors in the magnitude space normalized to the standard deviation of 1 and the mean of 0. We assume correlations only within Rubin or Roman bands, not between Rubin and Roman bands, since we have confirmed that the flux errors are not meaningfully correlated between Roman and Rubin bands. Therefore, the resultant local correlation matrices have a dimension of $3\times3$ ($6\times6$) for Roman (Rubin) bands. Note that because of this setting, each time we add a galaxy for testing, it has to go through the local correlation matrix calculation with the ``training'' sample. 
    \item To remove unphysical behaviors in the local correlation matrices, $C$, we apply a small regularization factor, $\epsilon=0.001$, to remove negative or extremely small eigenvalues:
    \begin{equation}
        C_{\rm reg}(X) = (1-\epsilon)C(X) + \epsilon I,
    \end{equation}
    where $I$ is an identity matrix. This ensures a positive-definite correlation matrix.\footnote{See, e.g., \citet{Alsing2024} and \citet{Fischbacher2025} for studies that use somewhat similar methods to ours.}
\end{itemize}

As a result, the final trained model consists of: 1) per-band conditional bias models, $\hat{B}_b(X)$, 2) per-band conditional residual (of fractional flux bias with respect to the median) distribution models, $R_{i,b} = \hat{Q}_{R_b}(u_{i,b} | X_i)$, and 3) a (true) magnitude-dependent local Student-\textit{t} correlation matrix, $C(X)$, with fixed Student-\textit{t} degrees of freedom $\nu$=4. 

We apply this model to the testing sample and to the new \texttt{Diffsky} mock catalog (Sec.~\ref{subsec:diffsky}) as follows:
\begin{itemize}
    \item We predict the median (bias) of fractional flux error given true galaxy properties using the trained model, $\hat{B}_b(X)$.
    \item For each galaxy, we predict the residual quantiles using the aforementioned 23 quantile values, \{$\hat{Q}_{R_b}(u_j|X)$\} ($j$ runs over the 23 quantile values).
    \item Given the true magnitudes of each galaxy \{$m_i$\} ($i$ runs over the bands within Rubin or Roman), standardized using the training sample's means and standard deviations, we retrieve the local correlation matrix in the $Z$-space using the nearest 500 neighbors in the standardized magnitude space.
    \item A multi-variate Student-\textit{t} distribution is fully determined by a correlation matrix $\boldsymbol{C}$ and a degree of freedom, $\nu$. A particular point drawn from this distribution, $\boldsymbol{Z}_{\rm draw} =\{Z_b\}$ is then specified by a random multi-variate Gaussian vector, $\boldsymbol{G}$, given the correlation matrix, and a random realization, $W$, from the $\chi^2_\nu$ distribution, such that:
    \begin{equation}
        \boldsymbol{Z}_{\rm draw} = \frac{\boldsymbol{G}}{\sqrt{W/\nu}}.
    \end{equation}
    \item We convert this latent draw, $\boldsymbol{Z}_{\rm draw}$, back into percentiles:
    \begin{equation}
        u_{b,{\rm draw}} = T_\nu (Z_{i,b, {\rm draw}}),
    \end{equation}
    then convert $u_{b,{\rm draw}}$ back into physical residuals, $R_{b,{\rm draw}}$, by interpolating the calculated \{$\hat{Q}_{R_b}(u_j|X)$\} above:
    \begin{equation}
        R_{b,{\rm draw}} = \hat{Q}_{R_b}(u_{b,{\rm draw}} | X_i).
    \end{equation}
    \item Finally, we restore the median bias of $R_b$ (calculated in the first step) to get the final predictions for the fractional flux residuals, $Y_{b,{\rm draw}}$:
    \begin{equation}
    \label{eq:Y_model}
        Y_{b,{\rm draw}} = R_{b,{\rm draw}} + \hat{B}_b(X_i).
    \end{equation}
\end{itemize}

In Fig.~\ref{fig:diagram_algorithm}, we show our algorithm to train the measured flux errors as a function of true galaxy properties and how we apply this model to a different mock galaxy catalog.


\bsp	
\label{lastpage}
\end{document}